\RequirePackage{ifpdf}
\ifpdf % We are running pdfTeX in pdf mode
\documentclass[pdftex]{sigma}
\else
\documentclass{sigma}
\fi

\numberwithin{equation}{section} \numberwithin{lemma}{section}
\numberwithin{remark}{section} \numberwithin{proposition}{section}
\numberwithin{theorem}{section}

\begin{document}

%\newcounter{mo}
%\newcommand{\mo}[1]
%{\stepcounter{mo}$^{\bf MO\themo}$%
%\footnotetext{\hspace{-3.7mm}$^{\blacksquare\!\blacksquare}$ {\bf MO\themo:~}#1}}

%\newcounter{bk}
%\newcommand{\bk}[1]
%{\stepcounter{bk}$^{\bf BK\thebk}$%
%\footnotetext{\hspace{-3.7mm}$^{\blacksquare\!\blacksquare}$ {\bf
%BK\thebk:~}#1}}

%%%%%%%%%%%%%%%%%%%%%%%%%%%%%%%%%%%%

\newcommand{\Si}{\Sigma}
\newcommand{\tr}{{\rm tr}}
\newcommand{\ad}{{\rm ad}}
\newcommand{\Ad}{{\rm Ad}}
\newcommand{\ti}[1]{\tilde{#1}}
\newcommand{\om}{\omega}
\newcommand{\Om}{\Omega}
\newcommand{\de}{\delta}
\newcommand{\al}{\alpha}
\newcommand{\te}{\theta}
\newcommand{\vth}{\vartheta}
\newcommand{\be}{\beta}
\newcommand{\la}{\lambda}
\newcommand{\La}{\Lambda}
\newcommand{\Ga}{\Gamma}
\newcommand{\D}{\Delta}
\newcommand{\ve}{\varepsilon}
\newcommand{\ep}{\epsilon}
\newcommand{\vf}{\varphi}
\newcommand{\vfh}{\varphi^\hbar}
\newcommand{\vfe}{\varphi^\eta}
\newcommand{\fh}{\phi^\hbar}
\newcommand{\fe}{\phi^\eta}
\newcommand{\G}{\Gamma}
\newcommand{\ka}{\kappa}
\newcommand{\ip}{\hat{\upsilon}}
\newcommand{\Ip}{\hat{\Upsilon}}
\newcommand{\ga}{\gamma}
\newcommand{\ze}{\zeta}
\newcommand{\si}{\sigma}

\def\hS{{\hat{S}}}

\newcommand{\li}{\lim_{n\rightarrow \infty}}
\def\mapright#1{\smash{
\mathop{\longrightarrow}\limits^{#1}}}

\newcommand{\mat}[4]{\left(\begin{array}{cc}{#1}&{#2}\\{#3}&{#4}
\end{array}\right)}
\newcommand{\thmat}[9]{\left(
\begin{array}{ccc}{#1}&{#2}&{#3}\\{#4}&{#5}&{#6}\\
{#7}&{#8}&{#9}
\end{array}\right)}
\newcommand{\beq}[1]{\begin{equation}\label{#1}}
\newcommand{\eq}{\end{equation}}
\newcommand{\beqn}[1]{\begin{eqnarray}\label{#1}}
\newcommand{\eqn}{\end{eqnarray}}
\newcommand{\p}{\partial}
\def\sq2{\sqrt{2}}
\newcommand{\di}{{\rm diag}}
\newcommand{\oh}{\frac{1}{2}}
\newcommand{\su}{{\bf su_2}}
\newcommand{\uo}{{\bf u_1}}
\newcommand{\SL}{{\rm SL}(2,{\mathbb C})}
\newcommand{\GLN}{{\rm GL}(N,{\mathbb C})}
\def\sln{{\rm sl}(N, {\mathbb C})}
\def\slk{{\rm sl}(k, {\mathbb C})}
\def\sl2{{\rm sl}(2, {\mathbb C})}
\def\SLN{{\rm SL}(N, {\mathbb C})}
\def\SLT{{\rm SL}(2, {\mathbb C})}
\def\PSLN{{\rm PSL}(N, {\mathbb C})}
\newcommand{\PGLN}{{\rm PGL}(N,{\mathbb C})}
\newcommand{\gln}{{\rm gl}(N, {\mathbb C})}
\newcommand{\PSL}{{\rm PSL}_2( {\mathbb Z})}
\def\f1#1{\frac{1}{#1}}
\def\lb{\lfloor}
\def\rb{\rfloor}
\def\sn{{\rm sn}}
\def\cn{{\rm cn}}
\def\dn{{\rm dn}}
\newcommand{\rar}{\rightarrow}
\newcommand{\upar}{\uparrow}
\newcommand{\sm}{\setminus}
\newcommand{\ms}{\mapsto}
\newcommand{\bp}{\bar{\partial}}
\newcommand{\bz}{\bar{z}}
\newcommand{\bw}{\bar{w}}
\newcommand{\bA}{\bar{A}}
\newcommand{\bG}{\bar{G}}
\newcommand{\bL}{\bar{L}}
\newcommand{\btau}{\bar{\tau}}

\newcommand{\tie}{\tilde{e}}
\newcommand{\tial}{\tilde{\alpha}}
\newcommand{\tiH}{\tilde{H}}
\newcommand{\tiM}{\tilde{M}}
\newcommand{\titiM}{\tilde{\tilde{M}}}
\newcommand{\tilt}{\tilde{t}}

\newcommand{\Sh}{\hat{S}}
\newcommand{\vtb}{\theta_{2}}
\newcommand{\vtc}{\theta_{3}}
\newcommand{\vtd}{\theta_{4}}

\def\mC{{\mathbb C}}
\def\mZ{{\mathbb Z}}
\def\mR{{\mathbb R}}
\def\mN{{\mathbb N}}

\def\frak{\mathfrak}
\def\gg{{\frak g}}
\def\gJ{{\frak J}}
\def\gS{{\frak S}}
\def\gL{{\frak L}}
\def\gG{{\frak G}}
\def\gk{{\frak k}}
\def\gK{{\frak K}}
\def\gl{{\frak l}}
\def\gh{{\frak h}}
\def\gH{{\frak H}}
\def\gt{{\frak t}}
\def\gT{{\frak T}}

\def\baal{\bar{\al}}
\def\babe{\bar{\be}}

\def\bfa{{\bf a}}
\def\bfb{{\bf b}}
\def\bfc{{\bf c}}
\def\bfd{{\bf d}}
\def\bfe{{\bf e}}
\def\bff{{\bf f}}
\def\bfg{{\bf g}}
\def\bfm{{\bf m}}
\def\bfn{{\bf n}}
\def\bfp{{\bf p}}
\def\bfu{{\bf u}}
\def\bfv{{\bf v}}
\def\bfr{{\bf r}}
\def\bfs{{\bf s}}
\def\bft{{\bf t}}
\def\bfx{{\bf x}}
\def\bfM{{\bf M}}
\def\bfR{{\bf R}}
\def\bfC{{\bf C}}
\def\bfS{{\bf S}}
\def\bfJ{{\bf J}}
\def\bfz{{\bf z}}
\def\bfnu{{\bf \nu}}
\def\bfsi{{\bf \sigma}}
\def\bfU{{\bf U}}

\def\clA{\mathcal{A}}
\def\clC{\mathcal{C}}
\def\clD{\mathcal{D}}
\def\clE{\mathcal{E}}
\def\clG{\mathcal{G}}
\def\clR{\mathcal{R}}
\def\clU{\mathcal{U}}
\def\clT{\mathcal{T}}
\def\clO{\mathcal{O}}
\def\clH{\mathcal{H}}
\def\clK{\mathcal{K}}
\def\clJ{\mathcal{J}}
\def\clI{\mathcal{I}}
\def\clL{\mathcal{L}}
\def\clM{\mathcal{M}}
\def\clN{\mathcal{N}}
\def\clQ{\mathcal{Q}}
\def\clW{\mathcal{W}}
\def\clZ{\mathcal{Z}}

\def\baf{{\bf f_4}}
\def\bae{{\bf e_6}}
\def\ble{{\bf e_7}}
\def\bag2{{\bf g_2}}
\def\bas8{{\bf so(8)}}
\def\baso{{\bf so(n)}}

\def\sr2{\sqrt{2}}
\newcommand{\ran}{\rangle}
\newcommand{\lan}{\langle}
\def\f1#1{\frac{1}{#1}}
\def\lb{\lfloor}
\def\rb{\rfloor}
\newcommand{\slim}[2]{\sum\limits_{#1}^{#2}}

\allowdisplaybreaks

\renewcommand{\PaperNumber}{067}

\FirstPageHeading

\ShortArticleName{1+1 Gaudin Model}

\ArticleName{1+1 Gaudin Model}

\Author{Andrei V.~ZOTOV}

\AuthorNameForHeading{A.V.~Zotov}

\Address{Institute of Theoretical and Experimental Physics, Moscow, Russia}
\Email{\href{mailto:zotov@itep.ru}{zotov@itep.ru}}

\ArticleDates{Received January 29, 2011, in f\/inal form July 03, 2011;  Published online July 13, 2011}

\Abstract{We study 1+1 f\/ield-generalizations of the rational and
elliptic Gaudin models. For ${\rm sl}(N)$ case we introduce equations
of motion and L-A pair with spectral parameter on the Riemann sphere
and elliptic curve. In ${\rm sl}(2)$ case we study the equations in
detail and f\/ind the corresponding Hamiltonian densities. The
$n$-site model describes $n$ interacting Landau--Lifshitz models of
magnets. The interaction depends on position of the sites (marked
points on the curve). We also analyze the 2-site case in its own
right and describe its relation to the principal chiral model. We
emphasize that 1+1 version impose a restriction on a choice of f\/lows
on the level of the corresponding 0+1 classical mechanics.}


\Classification{14H70; 33E05; 37K20; 37K10}

%\today
%\vfill\eject
%\tableofcontents \vfill\eject

\renewcommand{\thefootnote}{\arabic{footnote}}
\setcounter{footnote}{0}

\section{Introduction}

Gaudin model (or Gaudin magnet) was introduced by M.~Gaudin
\cite{Gaudin1} as a quasiclassical limit of
%XXX (and later XYZ)
spin-1/2 chain and was studied via the Bethe ansatz
\cite{Baxter1,SklTak}. Let us start with a general rational model
underlying Gaudin magnets. The classical rational Gaudin model is
def\/ined by the following quadratic Hamiltonians:
\begin{gather}\label{q01} H_a=-\frac{1}{2}\sum\limits_{c\neq a}\frac{\langle S^a
S^c\rangle}{z_a-z_c},\qquad  a=1,\dots,n, \end{gather} where $S^a\in\sl2$,
$\{z_1,\dots,z_n\}\in {\mathbb{CP}}^1$ are marked points and $\langle\
\ \rangle$ denotes the trace.

From the point of view of the Lax pair the model is described by a
general Lax matrix which is a $\sln$-valued function $L(z)$ on
${{\mathbb{CP}}^1\backslash\{z_1,\dots,z_n\}}$ with simple poles at
$\{z_1,\dots,z_n\}$ and some given residues
$\hbox{Res}_{z_a}L(z)=S^a\in\sln$:
\begin{gather}\label{q02} L(z)=\sum\limits_{a=1}^n\frac{S^a}{z-z_a}.
\end{gather}

The generating function of the Hamiltonians is
\begin{gather}\label{q2} \frac{1}{2}\langle L^2(z)\rangle
=\frac{1}{2}\sum\limits_{a=1}^n \frac{\langle
\left(S^a\right)^2\rangle}{(z-z_a)^2}+ \sum\limits_{a\neq
b}\frac{1}{z-z_a}\frac{\langle S^a S^b\rangle}{z_a-z_b}.
\end{gather}

The f\/irst sum in (\ref{q2}) shows that the eigenvalues $\la_a$ of
$S^a$ are the constant $\mathbb C$-numbers. Thus, the phase space is
a direct product\footnote{In fact, there is coadjoint action of
$\SLN$ on M which provides the constraint $\sum_a S^a=0$ with
some f\/ixation of $\SLN$ action. Then one can make a reduction
$M\rightarrow M//\SLN$. But we do not go into details of this
reduction here. In \cite{Zotov1,LZ} the examples of the reduction
for the Painlev{\'e} VI equation are discussed. The $r$-matrix of
the reduced models satisf\/ies the ref\/lection equations. Thus, the
models live on the boundaries of the f\/inite lattices.} of the
coadjoint orbits by ${{\rm SL}(N, {\mathbb C})}$ action:
$M={\mathcal O}^1\times$ $\dots\times{\mathcal O}^n$. This phase
space is naturally equipped with a linear Poisson--Lie structure:
\begin{gather}\label{q3}
\{S^a_\al,S^b_\be\}=\delta^{ab}\sum\limits_\ga
C_{\al\be}^\ga S^a_\ga,
\end{gather} where $S^a_\al$ are
coef\/f\/icients in some basis $\{T_\al\}$: $S^a=\sum\limits_{\al}
S^a_\al T_\al$ and $C_{\al\be}^\ga$ are the structure constants of
$\sln$ in this basis. The natural basis is described in the
appendix. The Hamiltonians (\ref{q01}) in $\sln$ case are replaced
by
\begin{gather}\label{q0101} H_a=-\frac{1}{N}\sum\limits_{c\neq a}\frac{\langle S^a
S^c\rangle}{z_a-z_c},\qquad  a=1,\dots,n. \end{gather}

The dynamics with respect to the Hamiltonians (\ref{q0101}) is given
by the following equations\footnote{Here we imply some choice of the
normalization by the Killing form $\langle\ \rangle$, see also
(\ref{Kil}).}:
\begin{gather*}
\p_{t_a}S^a=\{H_a,S^a\}=-\sum\limits_{c\neq a}\frac{[S^a,S^c]}{z_a-z_c},
\nonumber\\
\p_{t_a}S^b=\{H_a,S^b\}=\frac{[S^a,S^b]}{z_a-z_b} \qquad  \hbox{for}\ \ a\neq b.%\label{q5}
\end{gather*}

These equations of motion can be represented in the Lax form
\begin{gather}\label{q6} \p_{t_a}L=[L,M_a]
\end{gather}
with the Lax pair
\begin{gather*}%\label{q7}
 L(z)=\sum\limits_{c=1}^n\frac{S^c}{z-z_c}, \qquad
M_a(z)= \frac{S^a}{z-z_a}.
 \end{gather*}

In such a generality the model was studied many times. For example,
the non-autonomous version corresponds to the Schlesinger system of
the isomonodromic deformations on a sphere. It was studied a hundred
years ago~\cite{Schlez}.

In the elliptic case \cite{STSR} the Lax matrix (\ref{q02}) is
replaced by
\begin{gather}\label{q08}
L(z)=\sum\limits_{a=1}^n\sum\limits_{\al}S^a_\al\vf_\al(z-z_a)T_\al,
\end{gather} where $z\in\Sigma_\tau$ is a coordinate on an elliptic curve
$\Sigma_\tau$ with moduli $\tau$.  Basis $\{T_\al\}$ and the
corresponding Poisson structure is def\/ined in~(\ref{B.11}). Functions $\vf_\al(z-z_a)$ (\ref{vf}) form a basis in
$\Gamma(\hbox{End}\,V,\Sigma_\tau)$ with a simple pole at $z_a$ for
some f\/ixed holomorphic vector bundle~$V$ of degree one. The Poisson
structure (\ref{q3}) for the structure constants~(\ref{AA3b})
is related to the existence of the $r$-matrix of the
Belavin--Drinfel'd type~\cite{Belavin}. The quadratic Poisson
structure can be def\/ined by the same $r$-matrix~\cite{LOZCh}.

Most of problems natural for integrable systems have been studied
for the Gaudin model as well. Among them the separation of variables
\cite{Skl}, relations to monodromy preserving and
Knizhnik--Zamolodchikov equations~\cite{Skl2}, quantum quadratic
algebras and bihamiltonian structures \cite{KLO}, time-discrete
versions~\cite{discr}, quantization~\cite{RST} and Langlands duality
\cite{Langl}. It should be mentioned that the elliptic Gaudin model
was originally def\/ined by B.~Enriquez and V.~Rub\-tsov~\cite{ER} as an
example of the Hitchin-type system~\cite{H}. ``Dynamical'' case was
considered f\/irst by A.~Gorsky and N.~Nekrasov~\cite{GN}. That case
corresponded to degree zero vector bundle $V$ (that is to nontrivial
moduli space of bundles) or to the ``spin'' extensions of the Calogero
model. In~\cite{LOZ} it was shown that the top-like models and
Calogero-type models are related by means of the modif\/ication
procedure (the later changes the degree of~$V$). In this respect,
the models are equivalent.

Consideration of particular cases and dif\/ferent types of reductions
leads to relations between Gaudin model and a number of known
integrable systems such as interacting tops~\cite{Ragnisco},
Pain\-le\-v{\'e}~VI equation and Zhukovsky--Volterra gyrostat~\cite{Zotov1}, Neumann system~\cite{Kuznetsov}.

With the advent of the inverse scattering method the Lax equations
or the zero-curvature equations~\cite{ZSh} (with spectral parameter)
became a main tool for investigation of nonlinear equations~\cite{FT2,DKN}. Dif\/ferent applications and classif\/ications can be
found in~\cite{MS}. In this paper we are predominantly interested in
the Landau--Lifshitz equation~\cite{LL} (which describes the
continuous limit of the XYZ model~\cite{Sk,FT}) and the principal
chiral model~\cite{ZakhMikh,Chered}.

In \cite{LOZ} a general scheme was suggested for constructing $1+1$
(or f\/ield) generalizations of the Gaudin-type models as typical
examples of the Hitchin systems. As a by-product of this work the
f\/ield generalization of the elliptic Calogero model was
obtained\footnote{This result was f\/irst obtained by I.~Krichever in~\cite{K}.} and its equivalence to the Landau--Lifshitz equation was
shown in terms of the special singular gauge transformations.

{\it The purpose of the paper} is to present explicit L-A
pairs for $1+1$ Gaudin model, to propose corresponding Hamiltonian
description and to f\/ind out relationships between the obtained
equations and some known models such as the Heisenberg Model, the
Landau--Lifshitz equation and the principal chiral model.

In 1+1 models the Lax equations (\ref{q6}) are replaced by the
zero-curvature (or Zakharov--Shabat) equations:
\begin{gather*}%\label{q6007}
\p_{t_a}L-\p_x M_a=[L,M_a],
 \end{gather*}
  where $L$ and $M_a$ do
not coincide (in general) with those from~(\ref{q6}). It was shown
in \cite{LOZ} how to construct 1+1 version of $L$-operator. In
particular, $L$ keeps the same form as in (0+1) version of the
Hitchin systems corresponding to holomorphic vector bundles of
degree~1. This class of systems is under our consideration in this
paper. A general scheme~\cite{DMN} allows to obtain densities of the
conserved quantities (Hamiltonians). However, there is a technical
problem of f\/inding corresponding $M$-operators. Unfortunately, there
is no practical way to get them explicitly. For example, in
\cite{LOZ} the nontrivial $M$-operator for the f\/ield version of
Calogero model was obtained by some ansatz. In the same manner
$M$-operators were obtained in~\cite{Sk,Chered} for the
Landau--Lifshitz and the Principal Chiral Models correspondingly. The
inverse problem (to f\/ind mechanical~$L$ and~$M$ from known f\/ield
versions) is an easy task~-- one should put to zero all derivatives
with respect to the loop variable $x$. In this respect, there is a
correspondence between f\/ield f\/lows and some choice of f\/lows
($M$-operators) on the level of classical mechanics. It will be
shown that the f\/irst f\/lows of 1+1 Gaudin hierarchy correspond to
``conventional'' description of f\/lows in the Gaudin mechanics while
the second f\/lows arise naturally from some ``reformulated'' version.
The later appears as some linear combination of the ``conventional''
Gaudin f\/lows.

The paper is organized as follows: in Section~\ref{section2} we give
a standard description of the Gaudin model and its f\/lows $M_a$
(Proposition~\ref{proposition2.1}). Then the ``reformulated'' version is suggested in
the form of linear combinations of $\{M_a\}$ (Proposition~\ref{proposition2.2}). In
Section~\ref{section3} we discuss the f\/ield generalization and f\/ind the f\/irst
(Proposition~\ref{proposition3.1}) and the second (Proposition~\ref{proposition3.2}) f\/lows of the 1+1
Gaudin hierarchy. Among other things, we consider a special case of
the f\/irst f\/lows corresponding to the principal chiral model in
detail. In Section~\ref{section4} $\sl2$ case is considered (rational~--
Subsection~\ref{section4.1} and elliptic -- Subsection~\ref{section4.2}) and the Hamiltonian
description is obtained. First, we get general formulae for the
densities of Hamiltonians via local decomposition for the f\/irst
(Lemma~\ref{lemma4.1}) and the second (Lemma~\ref{lemma4.2}) f\/lows. Secondly, we evaluate
these densities for 1+1 rational and elliptic $\sl2$ Gaudin model
and reproduce previously obtained equations of motion (Theorem~\ref{theorem4.1}).

The results of the paper can be brief\/ly summarized as
follows:
\begin{gather*}%\label{q0801}
\begin{array}{@{}l}
\underline{\hbox{(0+1)}\ \hbox{mechanics:}}\\
\
\\
\hbox{Gaudin f\/lows}\ \{H_a\}
\\
\
\\
\
\\
\hbox{Gaudin f\/lows}\ \{\ti{H}_a\}
\\
\hbox{(``reformulated version'')}
\end{array}
\ \ \ \
\begin{array}{l}
\underline{\hbox{(1+1)}\ \hbox{f\/ield version:}}\\
\
\\
{\hbox{1}}^{\rm st}\ \hbox{f\/lows}\ \{\mathcal{H}_{a,1}\}
\\
\
\\
\
\\
{\hbox{2}}^{\rm nd}\ \hbox{f\/lows}\ \{\mathcal{H}_{a,2}\}
\\
\
\end{array}
\ \ \ \
\begin{array}{l}
\underline{\hbox{type of models:}}
\\
\
\\
\hbox{$n$-site generalization of}
\\
\hbox{principal chiral model}
\\
\
\\
\hbox{interacting models of}
\\
\hbox{Landau--Lifshitz type}
\end{array}
\end{gather*}

The f\/irst f\/lows are described by the following equations:
\begin{gather*}%\label{q03001}
\p_{{t}_a}S^a-k\p_x S^a=-\sum\limits_{c\neq
a}[S^a,\hat{\vf}_{ac}(S^c)],
\\
\p_{{t}_a}S^b=[S^b,\hat{\vf}_{ba}(S^a)].
\end{gather*}

In ``2-site'' case and rational limit these are the equations of the
principal chiral model:
\begin{gather*}%\label{q0600501}
  \p_tl_1-k\p_x l_0+\frac{2}{z_1-z_2}[l_1,l_0]=0,\\
\p_tl_0-k\p_x l_1=0
\end{gather*}
with $l_0=S^1+S^2$ and $l_1=S^1-S^2$.

The equations for the second f\/lows are of the form (here we put
$\sl2$ case and $\sln$ is considered below):
\begin{gather}
\p_{\tilt_a} S^a-k\p_x
\eta^a=[S^a,\hat{\wp}(S^a)]+\sum\limits_{c\neq
a}[\eta^a,\hat{\vf}_{ca}(S^c)]-[\hat{F}_{ca}(S^c),S^a],\nonumber\\
\p_{\tilt_a}
S^b=[\hat{\vf}_{ab}(\eta^a),S^b]+[S^b,\hat{F}_{ba}(S^a)],\label{q18101}
\end{gather}
where $\eta^a=-\frac{k}{4\la_a^2}[S^a,S^a_x]+\sum\limits_{c\neq
a}\hat{\vf}_{ac}(S^c)$. Note that in ``1-site'' case $n=1$ the f\/irst
one equation in (\ref{q18101}) is the Landau--Lifshitz equation (for
$t_1=t$):
\begin{gather*}%\label{q19101}
\p_t S+\frac{k^2}{4\la^2}[S,S_{xx}]=[S,\hat{\wp}(S)].
\end{gather*}

\section[$\sln$ elliptic Gaudin model]{$\boldsymbol{\sln}$ elliptic Gaudin model}\label{section2}

\subsection{Standard description}\label{section2.1}

The phase space of the Gaudin model is a direct product of  orbits
${\mathcal O}_1\times\dots\times {\mathcal O}_n$ by the coadjoint
action of $\SLN$).
The coordinates $\{S^c_\al\}$ on each orbit $S^c\in {\mathcal O}_c$
are chosen to be dual to the basis $\{T_\al\}$  of the Lie algebra
$\sln$. The later basis $\{T_\al\}$ is built as the projective
representation of $(\mZ/N\mZ\oplus\mZ/N\mZ)$ in $\GLN$ (see~(\ref{B.11})). The corresponding structure constants~(\ref{AA3b})
provides the Poisson--Lie brackets:
\begin{gather}\label{q001}
\{S^a_\al,S^b_\be\}=\delta^{ab}c_{\al,\be}S_{\al+\be}.
\end{gather}

Let us introduce now the Lax matrix def\/ined on the elliptic curve
$\Si_\tau={\mathbb C}/(\mZ+\tau\mZ)$ with modular parameter $\tau$
(${\rm Im}(\tau)>0$):
\begin{gather}\label{q002} L(z)=\sum\limits_{c=1}^n\sum\limits_{\al\in\Ga '} S^c_\al
T_\al\vf_\al(z-z_c), \end{gather} where  $\Ga_N '=\ti{\mZ}^{(2)}_N$
(see (\ref{B.10})) and functions $\{\vf_\al(z-z_c)\}$ form the basis
in the space of sections $\Ga(\hbox{End}\,V,\Si_\tau)$ with simple
poles at $\{z_c\},\ c=1,\dots,n$ for the holomorphic vector bundle $V$
of degree one associated with the principle $\GLN$-bundle over
$\Si_\tau$. In fact the Lax matrix is f\/ixed by the quasiperiodic
properties with~(\ref{q}),~(\ref{la}):
\begin{gather*}%\label{q00201}
L(z+1)=Q L(z)Q^{-1},\qquad  L(z+\tau)=\La
L(z)\La^{-1}\end{gather*} and residues $\hbox{Res}_{z_a}L(z)=S^a$.

The invariants of the Lax matrix generate commuting Hamiltonians\footnote{Note
that we use both the Eisenstein and the Weierstrass functions. They
are simply related (\ref{ax100}), (\ref{ax101}).}
\begin{gather*}%\label{q003}
\frac{1}{2N}\langle
L^2(z)\rangle=\sum\limits_{c=1}^n
\left(H_{2,c}\wp(z-z_c)-H_{1,c}E_1(z-z_c)\right)-H_0,
\end{gather*}
where
$H_{2,c}=\frac{1}{2N}\langle
\left(S^c\right)^2\rangle=\frac{1}{2}\sum\limits_{\al\in\Ga_N'}S^c_\al
S^c_{-\al}$ are the Casimir functions corresponding to the orbits~${\mathcal O}_c$ and the Hamiltonians are:
\begin{gather}\label{q004}
H_{1,a}=-\frac{1}{N}\sum\limits_{c\neq a}\langle S^a
{\hat\vf}_{ac}(S^c)\rangle= -\sum\limits_{c\neq
a}\sum\limits_{\al\in\Ga_N'} S^a_{-\al}S^c_\al\vf_\al(z_a-z_c),\\
 H_0=\frac{1}{2N}\sum\limits_{c}\langle S^c {\hat
\wp}(S^c)\rangle-\frac{1}{2N}\sum\limits_{b\neq c}\langle S^b {\hat
f}_{bc}(S^c)\rangle \nonumber\\
\phantom{H_0}{} =\frac{1}{2}\sum\limits_{ c}\sum\limits_{\al\in\Ga_N'}
S^c_{-\al}S^c_\al \wp(\om_\al)-\frac{1}{2}\sum\limits_{b\neq
c}\sum\limits_{\al\in\Ga_N'} S^b_{-\al}S^c_\al f_\al(z_b-z_c),\label{q005}
\end{gather}
where we use the following notations: $\wp(\om_\ga)$ is
def\/ined in (\ref{AA5}), functions $\vf_\ga(z)$ and $f_\ga(z)$  in~(\ref{vf}),~(\ref{f}). We also def\/ine the linear operators:
\begin{gather*}%\label{q006}
\hat\wp:\ S_\al\rightarrow S_\al\wp(\om_\al),\qquad
\hat\vf_{ab}:\ S_\al\rightarrow S_\al\vf_\al(z_a-z_b),\qquad \hat
f_{ab}:\ S_\al\rightarrow S_\al f_\al(z_a-z_b).
\end{gather*} In the
following we also use $\hat E_1:\ S_\al\rightarrow S_\al
E_1(\om_\al)$. Note that
\begin{gather}\label{q00602}
\hat\vf^*_{ab}=-\hat\vf_{ba}
\end{gather} in the
sense that $\langle S^a\hat\vf_{ab} (S^b)\rangle=-\langle
S^b\hat\vf_{ba} (S^a)\rangle$ due to (\ref{A.300}). Similarly, $\hat
f^*_{ab}=\hat f_{ba}$, $\hat\wp^*=\hat\wp$ and $\hat E_1^*=-\hat
E_1$.

The
commutativity of the Hamiltonians with respect to (\ref{q001}) follows from the
underlying linear $r$-matrix structure of the Belavin--Drinfel'd type:
$r_{12}^{\rm BD}(z,w)=\sum\limits_{\al\in\Ga '}\vf_\al(z-w)T_\al\otimes
T_{-\al}$ \cite{Belavin}. Note also that the Hamiltonians $H_{1,a}$ are not independent:
\begin{gather*}%\label{q00603}
\sum\limits_{a=1}^n H_{1,a}=-\frac{1}{N}\sum\limits_{a=1}^n\sum\limits_{c\neq a}\langle S^a
{\hat\vf}_{ac}(S^c)\rangle \; \overset{\eqref{q00602}}{=} \; 0.
\end{gather*}
The appropriate number of independent Hamiltonians is achieved by taking into account $H_0$ and all higher Hamiltonians.

Let us write down equations of motion with respect to the Hamiltonians (\ref{q004}), (\ref{q005}):
\begin{gather}\label{q007}
\p_{t_a}S^a=\{H_{1,a},S^a\}=-\sum\limits_{c\neq a}[S^a,\hat\vf_{ac}(S^c)],
\\
\label{q00701}
\p_{t_a}S^b=\{H_{1,a},S^b\}=[S^b,\hat\vf_{ba}(S^a)],
\\
\label{q008}
\p_{t_0}S^a=\{H_0,S^a\}=[S^a,\hat\wp (S^a)]-\sum\limits_{c\neq a}[S^a,\hat{f}_{ac}(S^c)].
\end{gather}

\begin{proposition}\label{proposition2.1}
The equations of motion \eqref{q007}--\eqref{q008} can be presented in the Lax form \eqref{q6}
with the Lax matrix $L(z)$ defined in \eqref{q002} and $M$-matrices
given as follows:
\begin{gather}\label{q009}
M_a=\sum\limits_{\al\in\Ga_N'}S^a_\al T_\al\vf_\al(z-z_a),
\\
\label{q010}
M_0=-\sum\limits_{b=1}^n\sum\limits_{\ga\in\Ga_N'}S^b_\ga T_\ga f_\ga(z-z_b).
\end{gather}
\end{proposition}

\begin{proof}
The proof is direct. It is based on the usage of (\ref{we4})--(\ref{we8}).

Let us prove identity (\ref{we8}) which is the most nontrivial here. For a generic point $w\in\Sigma_\tau$ consider
$m_\ga^a(z,w)=\vf_\ga(z-w)\vf_\ga(w-z_a)$:
\begin{gather*}
m_\ga^a \; \overset{\eqref{we7}}{=}\;\vf_\ga(z-z_a)(E_1(z-w)+E_1(w-z_a)+E_1(\om_\ga)-E_1(\om_\ga+z-z_a))
\\
\hphantom{m_\ga^a}{} \ \ =
\vf_\ga(z-z_a)(E_1(z-w)+E_1(w-z_a))-f_\ga(z-z_a).
\end{gather*}
Combining (\ref{we4}) and (\ref{we7}) which are implied to be known
we have:
\begin{gather*}
\vf_\be(z-z_c)m_\ga^a=(\vf_\be(z-z_c)\vf_\ga(z-w))\vf_\ga(w-z_a) \\
\quad{} \overset{\eqref{we4}}{=}
\vf_\be(w-z_c)(\vf_{\be+\ga}(z-w)\vf_\ga(w-z_a))+\vf_{\be+\ga}(z-z_c)(\vf_{\ga}(z_c-w)\vf_\ga(w-z_a))\\
\quad{} \overset{\eqref{we4}, \eqref{we7}}{=}
\vf_\be(w-z_c)\vf_{\be+\ga}(z-z_a)\vf_{-\be}(w-z_a)+\vf_\be(w-z_c)\vf_\ga(z-z_a)\vf_\be(z-w)
\\
\qquad {} +\vf_{\be+\ga}(z-z_c)\vf_\ga(z_c-z_a)(E_1(z_c-w)+E_1(w-z_a)+E_1(\om_\ga)-E_1(\om_\ga+z_c-z_a))\\
\quad {} \overset{\eqref{we7}}{=}
\vf_\ga(z-z_a)m_\be^c-
\vf_{\be+\ga}(z-z_a)\vf_\be(z_a-z_c)(E_1(w-z_c)+E_1(z_a-w)+E_1(\om_\be)\\
\qquad{} -E_1(\om_\be+z_a-z_c))
+ \vf_{\be+\ga}(z-z_c)\vf_\ga(z_c-z_a)(E_1(z_c-w)+E_1(w-z_a)+E_1(\om_\ga)\\
\qquad{} -E_1(\om_\ga+z_c-z_a))\\
\quad{}
= \vf_\ga(z-z_a)m_\be^c-\vf_{\be+\ga}(z-z_c)f_\ga(z_c-z_a)+\vf_{\be+\ga}(z-z_a)f_\be(z_a-z_c)
\\
\qquad{} +(E_1(z_c-w)+E_1(w-z_a))(\vf_{\be+\ga}(z-z_c)\vf_\ga(z_c-z_a)+\vf_{\be+\ga}(z-z_a)\vf_\be(z_a-z_c))\\
\quad{} \overset{\eqref{we4}}{=} \vf_\ga(z-z_a)m_\be^c-\vf_{\be+\ga}(z-z_c)f_\ga(z_c-z_a)+\vf_{\be+\ga}(z-z_a)f_\be(z_a-z_c)
\\
\qquad{} +(E_1(z_c-w)+E_1(w-z_a))\vf_{\ga}(z-z_a)\vf_\be(z-z_c).
\end{gather*}

This ends the proof of (\ref{we8}).
\end{proof}

\subsection{Useful reformulation}\label{section2.2}

In this subsection we rewrite the equations of motion in a form which
will be convenient for 1+1 generalization.
First, consider the following expressions for $a=1,\dots,n$:
\begin{gather}
\sum\limits_{\ga\in\Ga_N'}T_\ga\vf_\ga(z-z_a)\sum\limits_{c\neq a}S^c_\ga\vf_\ga(z_a-z_c)\nonumber\\
\quad {}
\overset{\eqref{we7}}{=} \sum\limits_{\ga\in\Ga_N'}T_\ga\sum\limits_{c\neq a}S^c_\ga\vf_\ga(z-z_c)
(E_1(z-z_a)+E_1(z_a-z_c)+E_1(\om_\ga)-E_1(z-z_c+\om_\ga))\nonumber\\
\quad {}= E_1(z-z_a)(L-M_a)+\sum\limits_{c\neq a} M_c E_1(z_a-z_c)+M_0+
\sum\limits_{\ga\in\Ga_N'}T_\ga S^a_\ga f_\ga(z-z_a)\nonumber\\
\quad{} =E_1(z-z_a)L+\sum\limits_{c\neq a} M_c E_1(z_a-z_c)+M_0-
\sum\limits_{\ga\in\Ga_N'}T_\ga S^a_\ga F_\ga(z-z_a).\label{q011}
\end{gather}
Then let us def\/ine new $M$-matrices in the following way:
\begin{gather*}%\label{q012}
\tilde{M}_a=\sum\limits_{\ga\in\Ga_N'}T_\ga S^a_\ga
F_\ga(z-z_a)+ \sum\limits_{\ga\in\Ga_N'}T_\ga
\eta'^a_\ga\vf_\ga(z-z_a),\qquad a=1,\dots,n,
\end{gather*} where
\begin{gather}\label{q013}
\eta'^a=\sum\limits_{c\neq a}T_\ga
S^c_\ga\vf_\ga(z_a-z_c)=\sum\limits_{c\neq a}M^c(z_a)=
\hbox{Res}_{z=z_a}\left(\frac{1}{z-z_a}L(z)\right).
\end{gather}
From
(\ref{q011}) we can see that the new $M$-matrices are the linear
combinations of (\ref{q009}), (\ref{q010}):
\begin{gather*}%\label{q014}
\tilde{M}_a=E_1(z-z_a)L+\sum\limits_{c\neq a} M_c E_1(z_a-z_c)+M_0.
\end{gather*}

Then the Lax equations yield
\begin{gather*}
\p_{\ti{t}_a}L=\left[L,\sum\limits_{c\neq a}M_cE_1(z_a-z_c)+M_0\right]=
\sum\limits_{c\neq a}E_1(z_a-z_c)\p_{t_c}L + \p_{t_0}L
\end{gather*}
and the equations of motion are:
\begin{gather*}
\p_{\ti{t}_a}S^a=\sum\limits_{c\neq a}E_1(z_a-z_c)\p_{t_c}S^a +\p_{t_0}S^a\\
\qquad{} \overset{\eqref{q007}-\eqref{q008}}{=}
\sum\limits_{c\neq a}[S^a, E_1(z_a-z_c)\hat\vf_{ac}(S^c)-\hat{f}_{ac}(S^c)]+[S^a,{\hat{\wp}}(S^a)]
\end{gather*} while for $b\neq a$:
\begin{gather*}
\p_{\ti{t}_a}S^b=\p_{t_b} S^b E_1(z_a-z_b)+
\sum\limits_{c\neq a,b}\p_{t_c}S^bE_1(z_a-z_c)+\p_{t_0}S^b=
[S^b,{\hat{\wp}} S^b]\\
\phantom{\p_{\ti{t}_a}S^b=}{} -E_1(z_a-z_b)\sum\limits_{c\neq b}[S^b,\hat\vf_{bc}(S^c)]
+\sum\limits_{c\neq a,b}E_1(z_a-z_c)[S^b,\hat\vf_{bc}(S^c)]-
\sum\limits_{c\neq b}[S^b,\hat{f}_{bc}(S^c)]\\
\phantom{\p_{\ti{t}_a}S^b=}{} =
[S^b,\hat\wp(S^b)+E_1(z_b-z_a)\hat\vf_{ba}(S^a)-\hat{f}_{ba}(S^a)]\\
\phantom{\p_{\ti{t}_a}S^b=}{}
+
\sum\limits_{c\neq a,b}[S^b,(E_1(z_b-z_a)+E_1(z_a-z_c))\hat\vf_{bc}(S^c)-\hat{f}_{bc}(S^c)].
\end{gather*} Finally, we have
\begin{gather*}%\label{q015}
\p_{\ti{t}_a}S^a=[S^a,\hat\wp S^a]+\sum\limits_{c\neq a}[S^a,\hat{F}_{ac}(S^c)],
\\
\p_{\ti{t}_a}S^b=[S^b,\hat\wp S^b]+[S^b,\hat{F}_{ba}(S^a)]+\sum\limits_{c\neq a,b}
[S^b,\hat\vf_{ba}(\hat\vf_{ac} (S^c))]\\
\phantom{\p_{\ti{t}_a}S^b}{}=\sum\limits_{c\neq a}[S^b,\hat\vf_{ba}(\hat\vf_{ac}
(S^c))]+[S^b,\hat{F}_{ba}(S^a)]=
[S^b,\hat\vf_{ba}(\eta'^a)]+[S^b,\hat{F}_{ba}(S^a)].%\label{q016}
\end{gather*}

The corresponding Hamiltonians are obtained in the same way:
\begin{gather}
H_0+\sum\limits_{c\neq a}E_1(z_a-z_c)H_c=\frac{1}{2N}\sum\limits_{c}\langle S^c\hat\wp S^c\rangle
-\frac{1}{2N}\sum\limits_{b\neq c}\langle S^c \hat{f}_{cb}(S^b)\rangle
\nonumber\\
\qquad{} -\frac{1}{N}\sum\limits_{c\neq a}E_1(z_a-z_c)
\sum\limits_{b\neq c}\langle S^c\hat\vf_{cb}(S^b)\rangle=
\frac{1}{2N}\sum\limits_{c}\langle S^c\hat\wp S^c\rangle
-\frac{1}{2N}\sum\limits_{b,c\neq a,\ b\neq c}\langle S^c \hat{f}_{cb}(S^b)\rangle
\nonumber\\
\qquad{}-\frac{1}{N}\sum\limits_{c\neq a}\langle S^a\hat{f}_{ac}(S^c)\rangle
-\frac{1}{2N}\sum\limits_{b,c\neq a,\ b\neq c}
(E_1(z_a-z_c)-E_1(z_a-z_b))\langle S^c \hat{\vf}_{cb}(S^b)\rangle
\nonumber\\
\qquad{}-\frac{1}{N}\sum\limits_{c\neq a}E_1(z_a-z_c)\langle S^c\hat\vf_{ca}(S^a)\rangle
\nonumber\\
\qquad{}=\frac{1}{2N}\sum\limits_{c}\langle S^c\hat\wp S^c\rangle+
\frac{1}{N}\sum\limits_{c\neq a}\langle S^a\hat{F}_{ac}(S^c)\rangle+
\frac{1}{2N}\sum\limits_{b,c\neq a,\ b\neq c}\langle S^c \hat\vf_{ca}(\hat\vf_{ab}(S^b))\rangle.\label{q017}
\end{gather}
The last one term equals:
\begin{gather}
\frac{1}{2N}\sum\limits_{b,c\neq a,\ b\neq c}\langle S^c \hat\vf_{ca}(\hat\vf_{ab}(S^b))\rangle=
\frac{1}{2N}\sum\limits_{b,c\neq a}\langle S^c \hat\vf_{ca}(\hat\vf_{ab}(S^b))\rangle-
\frac{1}{2N}\sum\limits_{c\neq a}\langle S^c \hat\vf_{ca}(\hat\vf_{ac}(S^c))\rangle
\nonumber\\
\qquad{}=\frac{1}{2N}\sum\limits_{b,c\neq a}\langle S^c \hat\vf_{ca}(\hat\vf_{ab}(S^b))\rangle-
\frac{1}{2N}\sum\limits_{c\neq a}\langle S^c \hat\wp(S^c)\rangle+
\frac{1}{2N}\sum\limits_{c\neq a}\langle S^c S^c\rangle\wp(z_a-z_c).\label{q018}
\end{gather} From (\ref{q017}), (\ref{q018}) we conclude that the Hamiltonians
for the reformulated version of the Gaudin model are of the form:
\begin{gather*}%\label{q019}
\ti{H}_a=\frac{1}{2N}\langle S^a\hat\wp S^a\rangle+
\frac{1}{N}\sum\limits_{c\neq a}\langle S^a\hat{F}_{ac}(S^c)\rangle+
\frac{1}{2N}\sum\limits_{b,c\neq a}\langle S^c
\hat\vf_{ca}(\hat\vf_{ab}(S^b))\rangle,\qquad  a=1,\dots,n
\end{gather*} or
\begin{gather*}
\ti{H}_a=\frac{1}{2N}\sum\limits_{c}\langle S^c\hat\wp S^c\rangle+
\frac{1}{N}\sum\limits_{c\neq a}\langle S^a\hat{F}_{ac}(S^c)\rangle+
\frac{1}{2N}\sum\limits_{\substack{b,c\neq a,\\ b\neq c}}\langle S^c
\hat\vf_{ca}(\hat\vf_{ab}(S^b))\rangle ,\qquad a=1,\dots,n.
\end{gather*}
Two last forms of the Hamiltonians are dif\/fer by  the constant
$\frac{1}{2N}\sum\limits_{c\neq a}\langle S^c
S^c\rangle\wp(z_a-z_c)$. Let us summarize the obtained in results in

\begin{proposition}\label{proposition2.2}
The dynamics of the Gaudin model produced by Hamiltonians
\begin{gather}\label{q020}
\ti{H}_a=\frac{1}{2N}\langle S^a\hat\wp S^a\rangle+
\frac{1}{N}\sum\limits_{c\neq a}\langle S^a\hat{F}_{ac}(S^c)\rangle+
\frac{1}{2N}\sum\limits_{b,c\neq a}\langle S^c \hat\vf_{ca}(\hat\vf_{ab}(S^b))\rangle
\end{gather}
is given by equations
\begin{gather*}%\label{q021}
\p_{\ti{t}_a}S^a=[S^a,\hat\wp S^a]+\sum\limits_{c\neq a}[S^a,\hat{F}_{ac}(S^c)],
\\
%\label{q022}
\p_{\ti{t}_a}S^b=[S^b,\hat\vf_{ba}(\eta'^a)]+[S^b,\hat{F}_{ba}(S^a)],\qquad  \eta'^a=\sum\limits_{c\neq a}\hat\vf_{ac}(S^c)
\end{gather*} and can be presented in the Lax form with $L(z)$ from \eqref{q002}
and \begin{gather*}%\label{q023}
\tilde{M}_a=\sum\limits_{\ga\in\Ga_N'}T_\ga S^a_\ga F_\ga(z-z_a)+
\sum\limits_{\ga\in\Ga_N'}T_\ga \eta'^a_\ga\vf_\ga(z-z_a),\qquad
a=1,\dots,n.
\end{gather*}
\end{proposition}

The Gaudin Hamiltonians (\ref{q004}) and (\ref{q020}) are simplif\/ied
when written in terms of $\eta'^a$ (\ref{q013}):
\begin{gather*}%\label{q024}
 H_{a}=-\langle S^a\eta'^a\rangle,\qquad
\ti{H}_a=\frac{1}{2N}\langle S^a\hat\wp
S^a\rangle-\frac{1}{2N}\langle \left(\eta'^a\right)^2\rangle+
\frac{1}{N}\sum\limits_{c\neq a}\langle S^a\hat{F}_{ac}(S^c)\rangle.
\end{gather*}

In the end of the section let us also give the rational
``reformulated'' version since it is more illuminating:
\begin{gather*}%\label{q025}
\tiM_a=\sum\limits_{c\neq a}\frac{M_c}{z_a-z_c}
+\frac{1}{z-z_a}L=\frac{1}{z-z_a}M_a+\frac{\eta'^a}{z-z_a},
\end{gather*}
where
\begin{gather}
\label{q026}
\eta'^a=\sum\limits_{c\neq a} \frac{S^c}{z_a-z_c}.
\end{gather}
Hamiltonians:
\begin{gather*}%\label{q0271}
\tiH_a=-\frac{1}{2N}\langle \left(\sum\limits_{c\neq a}
\frac{S^c}{z_a-z_c}\right)^2\rangle + \frac{1}{N}\sum\limits_{c\neq
a}\frac{\langle S^a S^c\rangle}{(z_a-z_c)^2}.
\end{gather*}
The later
follows from simple evaluation:
\begin{gather}
\sum\limits_{c\neq a}\frac{H_c}{z_a-z_c}=
-\frac{1}{N}\sum\limits_{c\neq a}\sum\limits_{b\neq
c}\frac{1}{z_a-z_c}\frac{\langle S^c S^b\rangle}{z_c-z_b}=
\frac{1}{N}\sum\limits_{c\neq a}\frac{\langle S^a
S^c\rangle}{(z_a-z_c)^2}\label{q027}\\
{}+\frac{1}{N}\sum\limits_{b,c\neq a;\ b\neq c}\frac{\langle S^c
S^b\rangle}{(z_c-z_a)(z_c-z_b)}=-\frac{1}{2N}\langle
\left(\eta'^a\right)^2\rangle + \frac{1}{N}\sum\limits_{c\neq
a}\frac{\langle S^a
S^c\rangle}{(z_a-z_c)^2}+\frac{1}{2N}\sum\limits_{c\neq
a}\frac{\langle \left(S^c\right)^2\rangle}{(z_a-z_c)^2}.\nonumber
\end{gather} The last one term is the analogue of the constant
$\frac{1}{2N}\sum\limits_{c\neq a}\langle S^c
S^c\rangle\wp(z_a-z_c)$ in (\ref{q018}). The corresponding equations
of motion are:
\begin{gather*}%\label{q028}
\p_{\tilt_a} S^a=\sum\limits_{c\neq
a}\frac{[S^a,S^c]}{(z_a-z_c)^2},\\
\p_{\tilt_a}
S^b=\frac{[S^b,S^a]}{(z_a-z_b)^2}+\frac{1}{z_a-z_b}\sum\limits_{c\neq
a}\frac{[S^c,S^b]}{z_a-z_c}=
\frac{[S^b,S^a]}{(z_a-z_b)^2}+\frac{[\eta'^a,S^b]}{z_a-z_b}.
 \end{gather*}

\section{Field version}\label{section3}

\subsection[1+1 $\sln$ Gaudin model]{1+1 $\boldsymbol{\sln}$ Gaudin model}\label{section3.1}

The general construction of the f\/ield version for the Hitchin
systems was described in \cite{LOZ}. For our current purposes we
only need to def\/ine the phase space. By analogy with mechanics let
us consider a collection (direct product) of $n$ orbits assigned to
the marked points, i.e.\ let $\hbox{Res}_{z=z_a}L(z)=S^a(x)$ be
elements of the loop coalgebras $\hat{\rm sl}^*(N, {\mathbb C})$ and
$x$ be a loop variable. We imply that the values of the invariants
under the coadjoint action (or the eigenvalues of $S^a$) are f\/ixed.
More over we assume for simplicity that the eigenvalues are $\mathbb
C$-numbers (independent of $x$). From the physical point of view it
means that the magnetic momentum vector is normalized (as it is
assumed in the Landau--Lifshitz model). The boundary conditions are
chosen to be periodic. In summary, $S^a(x)$ are ${\hat{\rm sl}^*(N,
{\mathbb C})}$-valued periodic functions on a unit circle ${\mathbb
S}^1$: $S^a(x+2\pi)=S^a(x)$ with eigenvalues $\{\la_{k,a},\
k=1,\dots,N,\ a=1,\dots,n\}$ f\/ixed to be $\mathbb{C}$-numbers:
$\p_x\la_a=0$.

In the f\/ield case the Lax equations (\ref{q6}) a replaced by the
zero-curvature equations:
\begin{gather}\label{q029} \p_{t_a} L-k\p_x M_a=[L,M_a].
\end{gather}

In fact, the numeration of $M_a$ should include two type of indices
as in (\ref{q004}): the f\/irst one type describes the number of the
f\/low in the hierarchy and runs over $1,\dots,N$ in 0+1 mechanics or
$1,\dots,\infty$ in 1+1 f\/ield theory while the second one runs over
$1,\dots,n$ in both cases and describes the assignment of the
Hamiltonians to the marked points. In this paper we are not going to
concern the whole hierarchy but only two f\/irst f\/lows (as we did in
0+1 case).

We will see that the f\/irst $n$ f\/lows of the hierarchy corresponds to
the Gaudin Hamiltonians in the standard description~(\ref{q004})
supplemented by the momenta~$P_a$ along $x$ while the second~$n$
f\/lows naturally related to reformulated version~(\ref{q020})
\begin{alignat*}{4}%\label{q0291}
&  \framebox{Standard description $H_a$} \ \ &&
 \longrightarrow \ \ && \framebox{$1^{\text{st}}$ f\/lows} & \\
& \framebox{Reformulated version $\ti{H}_a$}\ \ && \longrightarrow \ \ &&
\framebox{$2^{\text{nd}}$ f\/lows} & \\
& \hspace*{18mm}   \dots &&  \longrightarrow\ \ && \hspace*{4mm}  \dots &
  \end{alignat*}

Thus we do not use multi-index for times. It is suf\/f\/icient to use
$t_a$ and $\ti{t}_a$ for our purposes and we keep these notations
for the f\/ield version.

It should be mentioned that the f\/ield generalization of the Lax pair
into ``L-A'' pair satis\-fying~(\ref{q029}) is nontrivial. The fact that
the $L$-matrix (\ref{q002}) is unchanged in the f\/ield version
follows from the triviality of the moduli space of bundles of degree
one. It is explained in~\cite{LOZ} in detail. As a result we deal
with the following Lax matrix:
\begin{gather}\label{q0292}
L=\sum\limits_{c=1}^n\sum\limits_{\ga\in\Ga_N'}T_\ga
S^c_\ga \vf_\ga(z-z_c).
 \end{gather}

The $M_a$-matrices for the f\/irst f\/low coincide with the mechanical
versions either:
\begin{gather}\label{q0293}
M_a=\sum\limits_{\ga\in\Ga_N'}T_\ga S^a_\ga
\vf_\ga(z-z_a).
 \end{gather}

\begin{proposition}\label{proposition3.1}
The zero-curvature equations \eqref{q029} with $L$ from
\eqref{q0292} and $M_a$ from \eqref{q0293} are equivalent to the
following equations:
\begin{gather}
\p_{{t}_a}S^a-k\p_x S^a=-\sum\limits_{c\neq
a}[S^a,\hat{\vf}_{ac}(S^c)],\nonumber
\\
\p_{{t}_a}S^b=[S^b,\hat{\vf}_{ba}(S^a)].\label{q030}
 \end{gather}
 \end{proposition}

The proof is the same as in the 0+1 case. As we will see below the
Hamiltonian corresponding to $M_a$ has the form
\begin{gather*}%\label{q03017}
\mathcal{H}_a=\oint_{{\mathbb S}^1}\hbox{d}x\;
(P_a+H_a(S(x))),
\end{gather*} where $\oint_{{\mathbb
S}^1}\hbox{d}x\,P_a$ is the shift operator in the loop algebra
${\hat{\rm sl}(N, {\mathbb C})}$: $\{\oint_{{\mathbb
S}^1}\hbox{d}x\, P_a(x),S^b(y)\}=\delta_{ab}\p_yS^b(y)$ and $H_a$ is
def\/ined as in~(\ref{q0101}) or~(\ref{q004}). Thus the Hamiltonian
describing equations (\ref{q030}) has the form:
\begin{gather*}%\label{q05004}
\mathcal{H}_a=\oint_{{\mathbb
S}^1}\hbox{d}x  \left(P_a-\frac{1}{N}\sum\limits_{c\neq a}{\langle
S^a \hat\vf_{ac}(S^c)\rangle}\right).
\end{gather*}
The phase space is
a direct product of the symplectic orbits of the loop group
${\hat{\rm SL}(N, {\mathbb C})}$ with linear Poisson structure:
\begin{gather*}%\label{q050041}
\{S^a_\al(x),S^b_\be(y)\}=\delta_{ab}\delta(x-y)c_{\al,\be}S^a_{\al+\be}(x),\qquad a,b=1,\dots,n.
\end{gather*}

The second f\/lows are of our main interest.

\begin{proposition}\label{proposition3.2}
The zero-curvature equations
\begin{gather*}
\p_{\ti{t}_a} L-k\p_x \ti{M}_a=[L,\ti{M}_a] \end{gather*} with $L$
from \eqref{q0292} and
\begin{gather}\label{q031}
\tilde{M}_a=\sum\limits_{\ga\in\Ga_N'}T_\ga S^a_\ga
F_\ga(z-z_a)+ \sum\limits_{\ga\in\Ga_N'}T_\ga
\eta^a_\ga\vf_\ga(z-z_a),\qquad a=1,\dots,n, \end{gather} where
$\eta^a=\eta'^a+\Delta\eta^a$, $\eta'^a=\sum\limits_{c\neq
a}\hat\vf_{ac}(S^c)$ are equivalent to the following equations:
\begin{gather}
\p_{\ti{t}_a}S^a-k\p_x\eta^a=[S^a,\hat\wp(
S^a)]+\sum\limits_{c\neq
a}[S^a,\hat{F}_{ac}(S^c)]+\sum\limits_{c\neq
a}[\hat{\vf}_{ac}(S^c),\eta^a]
\nonumber\\
\phantom{\p_{\ti{t}_a}S^a-k\p_x\eta^a=} {}+\big[\hat{E}_1(S^a),\Delta\eta^a\big]+\big[S^a,\hat{E}_1(\Delta\eta^a)\big]
-\hat{E}_1\left[S^a,\Delta\eta^a\right],
\nonumber\\
\p_{\ti{t}_a}S^b=[S^b,\hat\vf_{ba}(\eta^a)]+[S^b,\hat{F}_{ba}(S^a)],
\nonumber\\
-k\p_x S^a=[S^a,\Delta\eta^a].\label{q032}
 \end{gather}
\end{proposition}

The proof is also similar to the one given for the 0+1 case.
Functions $\eta^a$ are not uniquely def\/ined by equations $-k\p_x
S^a=[S^a,\Delta\eta^a]$. We f\/ix this ambiguity by requiring
$\eta^a\rightarrow\eta'^a=\sum\limits_{c\neq a}\hat\vf_{ac}(S^c)$ or
$\Delta\eta^a\rightarrow 0$ in $0+1$ limit. As for the equation
$-k\p_x S^a=[S^a,\Delta\eta^a]$ itself only some  special cases were
studied such as ``vector'' case \cite{GolSok, KrichVol} and
``Grassmannian'' case (special coadjoint orbits) \cite{Skryp}. For
${\hat{\rm sl}(2, {\mathbb C})}$ case the answer is well known:
$\Delta\eta^a=-\frac{k}{4\la_a^2}[S^a,S^a_x]$.

\subsection{2-site case and principal chiral model}\label{section3.2}

L-A pair for the principal chiral model was suggested in
\cite{ZakhMikh} (see also \cite{FT2,Chered,Holod,Orlov}). Consider
the f\/irst f\/lows of the Gaudin model (\ref{q030}) with 2 sites or
marked points ($n=2$). It is convenient to start from the rational
version:
\begin{gather*}%\label{q06001}
 L=\frac{S^1}{z-z_1}+\frac{S^2}{z-z_2}=M_1+M_2.  \end{gather*}
 The
corresponding $M$-matrix is known to be
\begin{gather*}%\label{q06002}
M=M_1-M_2=\frac{S^1}{z-z_1}-\frac{S^2}{z-z_2}.
\end{gather*}
Therefore
the equations of motion are
\begin{gather}   \p_t S^1-k\p_x S^1=-\frac{2}{z_1-z_2}[S^1,S^2],\nonumber\\
\p_t S^2+k\p_x S^2=\frac{2}{z_1-z_2}[S^1,S^2].\label{q06003}
\end{gather}
Then the Hamiltonian describing equations (\ref{q06003}) has a form\footnote{See Section~\ref{section4.3} and (\ref{q30}).}:
\begin{gather*}%\label{q06004}
\mathcal{H}=\mathcal{H}_1-\mathcal{H}_2=\oint_{{\mathbb
S}^1}\hbox{d}x  \left(P_1-P_2-\frac{\langle S^1
S^2\rangle}{z_1-z_2}\right)
\end{gather*} and the phase space is a
direct product of two symplectic orbits of the loop group ${\hat{\rm
SL}(N, {\mathbb C})}$ with the linear Poisson structure:
\begin{gather*}%\label{q060041}
\{S^a_\al(x),S^b_\be(y)\}=\delta_{ab}\delta(x-y)c_{\al,\be}S^a_{\al+\be}(x),\qquad  a,b=1,2.
\end{gather*}

\begin{remark}\label{remark3.1}
One can make a substitution
$S^1=\frac{1}{2}(l_0+l_1)$ and $S^2=\frac{1}{2}(l_0-l_1)$ to
represent equations (\ref{q06003}) in its traditional
form
\begin{gather*}%\label{q06005}
 \p_tl_1-k\p_x l_0+\frac{2}{z_1-z_2}[l_1,l_0]=0,\\
\p_tl_0-k\p_xl_1=0
  \end{gather*}
\emph{or change the coordinates $(x,t)$ to ``light-cone'' coordinates
$\xi=\frac{t+k^{-1}x}{2}$, $\eta=\frac{t-k^{-1}x}{2}$:}
\begin{gather}
 \p_\eta S^1=-\frac{2}{z_1-z_2}[S^1,S^2],\nonumber\\
\p_\xi S^2=\frac{2}{z_1-z_2}[S^1,S^2].\label{q06006}
\end{gather}
\end{remark}

{\bf Elliptic case.} For L-A pair $L=M_1+M_2$ and $M=M_1-M_2$
with $M_a=\sum\limits_{\al\in\Gamma'_N}T_\al S^a_\al\vf_\al(z-z_a)$,
$a=1,2$ the equations (\ref{q030}) yields
($\p_t=\p_{t_1}-\p_{t_2}$):
\begin{gather*}%\label{q06007}
  \p_t S^1-k\p_x S^1=-2[S^1,\hat\vf_{12}(S^2)],\\
\p_t S^2+k\p_x S^2=2[S^2,\hat\vf_{21}(S^1)].
  \end{gather*} or by analogy with (\ref{q06006})
\begin{gather*}%\label{q06008}
  \p_\eta S^1=-2[S^1,\hat\vf_{12}(S^2)],\\
\p_\xi S^2=2[S^2,\hat\vf_{21}(S^1)].
  \end{gather*}

In $\sl2$ case this result was obtained by I.~Cherednik
\cite{Chered}. Here we see that the principal chiral model
corresponds to the special (2-site) case of the f\/irst f\/lows of $1+1$
Gaudin model. It should be also mentioned that in \cite{Chered} the
equations for $\sl2$ case were obtained as a f\/ield version of XYZ
model, i.e.\ from the second f\/low of 1-site Gaudin model (or $\sl2$
elliptic top). It may be explained as follows: consider stationary
solutions $S^a=S^a(\eta)$ (or $\p_\xi S^a=0$). Then f\/ixing the
ambiguity in solutions of the equation $[S^2,\hat\vf_{21}(S^1)]=0$
as $S^2=-\frac{1}{2}\hat\vf_{21}(S^1)$ we have
\begin{gather*}
\p_\eta S^1=[S^1,\hat\vf_{12}\hat\vf_{21}(S^1)]=[S^1,\hat\wp(S^1)],
\end{gather*}
which is the equation of $\sln$ elliptic top (or 1-site
elliptic Gaudin model) corresponding to the second f\/low
$H=\frac{1}{2N}\langle S^1\hat\wp(S^1)\rangle$.

\section[$\sl2$ 1+1 Gaudin models]{$\boldsymbol{\sl2}$ 1+1 Gaudin models}\label{section4}

\subsection{1+1 XXX Gaudin magnet: interacting Heisenberg models}\label{section4.1}

Let us consider the case $\hbox{Res}_{z_a}L(z)=S^a\in\sl2$ in
detail. The linear Poisson--Lie structure in this case:
\begin{gather*}%\label{q3'}
\{S^a_\al,S^b_\be\}=2\sqrt{-1}\delta^{ab}\ve_{\al\be\ga}S^a_\ga,
\end{gather*} where $S^a_\al$ are coef\/f\/icients in the basis of Pauli
matrices: $S^a=\sum\limits_{\al=1}^3 S^a_\al\si_\al$.

The Gaudin Hamiltonians are:
\begin{gather}\label{q315}
H_a=-\sum\limits_{c\neq
a}\frac{S^a_1S^c_1+S^a_2S^c_2+S^a_3S^c_3}{z_a-z_c},
 \end{gather}
while the Hamiltonians of the reformulated version are
\begin{gather}\label{q316}
\tiH_a= \frac{1}{2}\sum\limits_{c\neq a}\frac{\langle S^a
S^c\rangle}{(z_a-z_c)^2}-\frac{1}{4}\langle\left(\sum\limits_{c\neq
a}\frac{S^c}{z_a-z_c}\right)^2\rangle+\frac{H_a^2}{2\la_a^2}.
\end{gather}
Since \begin{gather} \sum\limits_{c\neq
a}\frac{H_c}{z_a-z_c}+\frac{H_a^2}{2\la_a^2}=
-\frac{1}{2}\sum\limits_{c\neq a}\sum\limits_{b\neq
c}\frac{1}{z_a-z_c}\frac{\langle S^c
S^b\rangle}{z_c-z_b}+\frac{H_a^2}{2\la_a^2}
\nonumber\\
\qquad{}=\frac{1}{2}\sum\limits_{c\neq
a}\frac{\langle S^a
S^c\rangle}{(z_a-z_c)^2}+\frac{1}{2}\sum\limits_{b,c\neq a;\ b\neq
c}\frac{\langle S^c
S^b\rangle}{(z_c-z_a)(z_c-z_b)}+\frac{H_a^2}{2\la_a^2}
\nonumber\\
\qquad{}=\frac{1}{2}\sum\limits_{c\neq a}\frac{\langle S^a
S^c\rangle}{(z_a-z_c)^2}-\frac{1}{4}\sum\limits_{b,c\neq a;\ b\neq
c}\frac{\langle S^c
S^b\rangle}{(z_a-z_c)(z_a-z_b)}+\frac{H_a^2}{2\la_a^2}
\nonumber\\
\qquad{}=\frac{1}{2}\sum\limits_{c\neq a}\frac{\langle S^a
S^c\rangle}{(z_a-z_c)^2}+\frac{1}{4}\sum\limits_{c\neq
a}\frac{\langle
\left(S^c\right)^2\rangle}{(z_a-z_c)^2}-\frac{1}{4}\langle\left(\sum\limits_{c\neq
a}\frac{S^c}{z_a-z_c}\right)^2\rangle+\frac{H_a^2}{2\la_a^2},\label{q81}
 \end{gather}
then the corresponding equations of motion are:
\begin{gather*}%\label{q9}
\p_{\tilt_a} S^a=\sum\limits_{c\neq
a}\frac{[S^a,S^c]}{(z_a-z_c)^2}-\frac{H_a}{\la_a^2}\sum\limits_{c\neq
a}\frac{[S^a,S^c]}{z_a-z_c},\\
\p_{\tilt_a}
S^b=\frac{[S^b,S^a]}{(z_a-z_b)^2}+\frac{1}{z_a-z_b}\sum\limits_{c\neq
a}\frac{[S^c,S^b]}{z_a-z_c}+
\frac{H_a}{\la_a^2}\frac{[S^a,S^b]}{z_a-z_b}=
\frac{[S^b,S^a]}{(z_a-z_b)^2}+\frac{[\eta'^a,S^b]}{z_a-z_b},
 \end{gather*}
where
\begin{gather}\label{q10}
\eta'^a=\sum\limits_{c\neq a}
\frac{S^c}{z_a-z_c}+\frac{H_a}{\la_a^2}S^a.
\end{gather}

\begin{remark}\label{remark4.1}
(\ref{q10}) dif\/fers from (\ref{q026}) by
$\frac{H_a}{\la_a^2}S^a$ and the corresponding Hamiltonian~(\ref{q81}) dif\/fers from (\ref{q027}) by $\frac{H_a^2}{2\la_a^2}$.
This dif\/ference does not follow from ansatz~(\ref{q031}) but appears
from the Hamiltonian description (see Section~\ref{section4.3}). The
corresponding Lax pair is given by $L$ from (\ref{q02}) and
\begin{gather*}%\label{q11}
\tiM_a=\sum\limits_{c\neq a}\frac{M_c}{z_a-z_c}
+\frac{1}{z-z_a}L+\frac{H_a}{\la_a^2}M_a=\frac{1}{z-z_a}M_a+\frac{\eta'^a}{z-z_a}.
\end{gather*}
\end{remark}

{\bf 1+1 version.} Let $S^a(x)\in {\widehat{\rm sl}(2, {\mathbb C})}$ be periodic
$\sl2$-valued functions on a circle ${\mathbb S}^1$:
$S^a(x+2\pi)=S^a(x)$ with eigenvalues $\{\la_a\}$ f\/ixed to be
$\mathbb{C}$-numbers: $\p_x\la_a=0$. The Poisson structure now is
\begin{gather}\label{q3'x}
\{S^a_\al(x),S^b_\be(y)\}=2\sqrt{-1}\delta^{ab}\ve_{\al\be\ga}S^a_\ga(x)\delta(x-y).
\end{gather}

Consider \begin{gather*}%\label{q12}
\tilde{M}_a=\frac{S^a}{(z-z_a)^2}+\frac{\eta^a}{z-z_a},
\end{gather*}
where
\begin{gather}\label{q13}
\eta^a=-\frac{k}{4\la_a^2}[S^a,S^a_x]+\sum\limits_{c\neq
a}
\frac{S^c}{z_a-z_c}+\frac{H_a}{\la_a^2}S^a=-\frac{k}{4\la_a^2}[S^a,S^a_x]+\eta'^a,\qquad
 S^a_x\equiv\p_xS^a.
 \end{gather}
 Let us remark here that in (0+1)
limit $\eta^a=\eta'^a$ (\ref{q10}). Then the zero-curvature equation
\begin{gather}\label{q14}
\p_{\tilt_a} L-k\p_x \tiM_a=[L,\tiM_a]
\end{gather} reads as
follows
\begin{gather}\label{q15}
\p_{\tilt_a} S^a-k\p_x \eta^a=\left[\sum\limits_{c\neq a}
\frac{S^c}{z_a-z_c},\eta^a\right]+\sum\limits_{c\neq
a}\frac{[S^a,S^c]}{(z_a-z_c)^2},\\
\p_{\tilt_a}
S^b=\frac{[S^b,S^a]}{(z_a-z_b)^2}+\frac{[\eta^a,S^b]}{z_a-z_b}.
\end{gather}

These equation generalize the Heisenberg model which appears from
(\ref{q15}) in $n=1$ (1-site) case:
\begin{gather*}%\label{q151}
\p_t S+\frac{k^2}{4\la^2}[S,S_{xx}]=0
\end{gather*} and described
by the Hamiltonian
\begin{gather*}%\label{q159}
\mathcal{H}=\frac{k^2}{16\la^2}\oint_{{\mathbb
S}^1}\hbox{d}x\; \langle\left(\p_xS\right)^2\rangle.
\end{gather*}

\subsection{1+1 XYZ Gaudin magnet: interacting Landau--Lifshitz models}\label{section4.2}

By analogy with the previous section the Hamiltonians in 0+1 $\sl2$
case:
\begin{gather}\label{q116}
\tiH_a= \frac{1}{4}\langle
S^a\hat\wp(S^a)\rangle+\frac{1}{2}\sum\limits_{c\neq a}\langle S^a
\hat{F}(S^c)\rangle-\frac{1}{4}\langle\left(\sum\limits_{c\neq
a}\hat\vf_{ac}(S^c)\right)^2\rangle+\frac{H_a^2}{2\la_a^2}.
\end{gather}

Consider now the following L-A pair:
\begin{gather*}%\label{q16}
 L(z)=\sum\limits_{c=1}^n\sum\limits_{\al=1}^3 S^c_\al\si_\al\vf_\al(z-z_c), \\
\tiM_a(z)= \sum\limits_{\al=1}^3
\eta^a_\al\si_\al\vf_\al(z-z_a)+S^a_\al\si_\al\vf_\be(z-z_a)\vf_\ga(z-z_a),
 \end{gather*} where $\al$, $\be$, $\ga$ are dif\/ferent indices equivalent
to 1, 2, 3 up to a cyclic permutation and (compare with (\ref{q13}))\footnote{The remark
about the term $\frac{H_a}{\la_a^2}S^a$ in the previous section is
reasonable here as well.}
\begin{gather}\label{q17}
\eta^a=-\frac{k}{4\la_a^2}[S^a,S^a_x]+\sum\limits_{c\neq
a}\hat{\vf}_{ac}(S^c)+\frac{H_a}{\la_a^2}S^a.
\end{gather}

The zero curvature equation (\ref{q14}) leads to equations of
motion:
\begin{gather}\label{q18}
\p_{\tilt_a} S^a-k\p_x
\eta^a=[S^a,\hat{\wp}(S^a)]+\sum\limits_{c\neq
a}[\eta^a,\hat{\vf}_{ca}(S^c)]-\hat{\vf}_{ca}([S^c,\hat{\vf}_{ca}(S^a)]),\nonumber\\
\p_{\tilt_a}
S^b=[\hat{\vf}_{ab}(\eta^a),S^b]+\hat{\vf}_{ba}([\hat{\vf}_{ba}(S^b),S^a])
\end{gather}
and $\eta^a$ (\ref{q17}) is a particular solution of the equation
\begin{gather*}%\label{q19}
-kS^a_x=\left[S^a,\eta^a-\sum\limits_{c\neq
a}\hat{\vf}_{ac}(S^c)\right].
\end{gather*}
It is f\/ixed if we require
$\eta^a\rightarrow \eta^a=\sum\limits_{c\neq
a}\hat{\vf}_{ac}(S^c)+\frac{H_a}{\la_a^2}S^a$ in (0+1) limit. The
proof follows from (\ref{a2})--(\ref{a5}). In particular, from
(\ref{a31}) it follows that
$\hat{\vf}_{ca}([S^c,\hat{\vf}_{ca}(S^a)])=[\hat{F}_{ca}(S^c),S^a]$.
Then~(\ref{q18}) is written in the form close to (\ref{q032}):
\begin{gather}
\p_{\tilt_a} S^a-k\p_x
\eta^a=[S^a,\hat{\wp}(S^a)]+\sum\limits_{c\neq
a}[\eta^a,\hat{\vf}_{ca}(S^c)]-[\hat{F}_{ca}(S^c),S^a],\nonumber\\
\p_{\tilt_a}
S^b=[\hat{\vf}_{ab}(\eta^a),S^b]+[S^b,\hat{F}_{ba}(S^a)].\label{q181}
 \end{gather}

The last three terms in the f\/irst equation of (\ref{q032}) vanish.
It is due to (\ref{a22}) that $\forall\, A,B\in\sl2$:
\begin{gather*}%\label{q181001}
\hat{E}_1([A,B])=[\hat{E}_1(A),B]+[A,\hat{E}_1(B)].
\end{gather*}

Note that in case $n=1$ the f\/irst one equation in (\ref{q181}) is
the Landau--Lifshitz equation:
\begin{gather*}%\label{q191}
\p_t S+\frac{k^2}{4\la^2}[S,S_{xx}]=[S,\hat{\wp}(S)]
\end{gather*}
described by the Hamiltonian
\begin{gather*}%\label{q192}
\mathcal{H}=\oint_{{\mathbb S}^1}\hbox{d}x
\left(\frac{1}{4}\langle
S\hat\wp(S)\rangle+\frac{k^2}{16\la^2}\langle\left(\p_xS\right)^2\rangle\right).
\end{gather*}

\subsection{Hamiltonian description}\label{section4.3}

The explicit form of the conserved quantities in terms of the f\/ields
are obtained by solving the Riccati equation. First, we make the
gauge transformation (see for example \cite{LOZ,DMN}):
\begin{gather*}%\label{q20}
\left(k\p_x+\mat{L_{11}}{L_{12}}{L_{21}}{-L_{11}}\right)\left(
\begin{array}{l}
\psi_1
\\
\psi_2
\end{array}
\right)=0 \rightarrow\left(k\p_x+\mat{0}{1}{T}{0}\right)\left(
\begin{array}{l}
\psi_1
\\
\psi_2
\end{array}
\right)=0 \end{gather*} with
\begin{gather}\label{q201}
 T=L_{12}L_{21}+L_{11}^2+kL_{11}\frac{\p_x
L_{12}}{L_{12}}-k\p_x L_{11}-\frac{k^2}{2}\frac{\p_x^2
L_{12}}{L_{12}}+\frac{3k^2}{4}\left(\frac{\p_x
L_{12}}{L_{12}}\right)^2 ,
\end{gather}
which leads to the
Schr{\"{o}}dinger equation:
\begin{gather*}%\label{q2199}
\big({-}k^2\p_x^2+T\big)\psi_1=0.
\end{gather*}

Taking wave function in the form
$\psi_1=e^{\frac{1}{k}\int_{x_0}^x {\rm d}y\; \chi(y)}$ we come to
the Riccati equation:
\begin{gather}\label{q22}
k\p_x\chi+\chi^2-T=0.
\end{gather}

The solution is obtained via local decompositions:
\begin{gather}\label{q23}
\chi_a=\frac{1}{z-z_a}\chi_{a,-1}+\chi_{a,0}+(z-z_a)\chi_{a,1}+\cdots,\\
T=\frac{1}{(z-z_a)^2}T_{a,-2}+\frac{1}{z-z_a}T_{a,-1}+T_{a,0}+\cdots,\\
\label{q231}
L=\frac{1}{z-z_a}L^{a,-1}+L^{a,0}+(z-z_a)L^{a,1}+\cdots .
\end{gather}

Then (\ref{q22}) gives:
\begin{gather}
\chi_{a,-1}^2\equiv\la_a^2=T_{a,-2},
\nonumber\\
\chi_{a,0}=\frac{1}{2\chi_{a,-1}}(T_{a,-1}-k\p_x\chi_{a,-1}),
\nonumber\\
\chi_{a,1}=\frac{1}{2\chi_{a,-1}}\left(T_{a,0}-\frac{T_{a,-1}^2}{4T_{a,-2}}\right),\label{q24}
\\
\cdots\cdots\cdots\cdots\cdots\cdots\cdots\cdots\cdots\cdots\nonumber
\end{gather}
As it was shown in \cite{DMN} $\chi_{a,k}$ (\ref{q24}) are the
densities of the conservation laws. We will use notation
\begin{gather*}%\label{q243}
h_{a,k}(x)=-\la_a\chi_{a,k-1}
\end{gather*}
for the densities and
\begin{gather*}%\label{q2439}
\mathcal{H}_{a,k}=\oint_{{\mathbb S}^1}\hbox{d}x\;
h_{a,k}(x)
\end{gather*}
for the Hamiltonians. The coef\/f\/icients
(\ref{q231}) of the decomposition of $L$-matrices in rational and
elliptic cases:
\begin{alignat}{3}
& L^{a,-1}=S^a, \qquad && L^{a,-1}=S^a, & \nonumber\\
& L^{a,0}=\sum\limits_{c\neq a}\frac{S^c}{z_a-z_c}, \qquad && L^{a,0}=\sum\limits_{c\neq a}\hat\vf_{ac}(S^c), &\nonumber\\
& L^{a,1}=-\sum\limits_{c\neq a}\frac{S^c}{(z_a-z_c)^2}, \qquad &&   L^{a,1}=-\frac{1}{2}\hat\wp(S^a)-\sum\limits_{c\neq
a}\hat{F}_{ac}(S^c), & \label{q21}\\
&\cdots\cdots\cdots\cdots\cdots\cdots \qquad && \cdots\cdots\cdots\cdots\cdots\cdots &\nonumber
\end{alignat}

In what follows we sometimes omit for the simplicity the index $a$
for $L$-matrix and its elements assuming decompositions
(\ref{q23})--(\ref{q231}). Substituting (\ref{q231}) into (\ref{q201})
we get: \begin{gather*}%\label{q25}
T_{a,-2}=L_{12}^{-1}L_{21}^{-1}+L_{11}^{-1}L_{11}^{-1},
\\
T_{a,-1}=L_{12}^{-1}L_{21}^{0}+L_{12}^{0}L_{21}^{-1}+2L_{11}^{0}L_{11}^{-1}+
k\frac{L_{11}^{-1}\p_x L_{12}^{-1}}{L_{12}^{-1}}-k\p_x L_{11}^{-1},
\\
T_{a,0}=L_{12}^{1}L_{21}^{-1}+L_{12}^{-1}L_{21}^{1}+2L_{11}^{1}L_{11}^{-1}+
L_{12}^{0}L_{21}^{0}+L_{11}^{0}L_{11}^{0} \\
\phantom{T_{a,0}=}{}+ \frac{k}{L_{12}^{-1}}\left(L_{11}^0\p_x L_{12}^{-1}+L_{11}^{-1}\p_x
L_{12}^0-\frac{L_{12}^0L_{11}^{-1}\p_x
L_{12}^{-1}}{L_{12}^{-1}}\right)-k\p_x L_{11}^0\\
\phantom{T_{a,0}=}{}
-\frac{k^2}{2}\frac{\p_x^2
L_{12}^{-1}}{L_{12}^{-1}}+\frac{3k^2}{4}\left(\frac{\p_x
L_{12}^{-1}}{L_{12}^{-1}}\right).
\end{gather*}

Let us summarize the obtained results.

\begin{lemma}\label{lemma4.1}
The density of the Hamiltonian $\chi_{a,0}$ has the following form
in terms of the decomposition of the $L$-matrix \eqref{q231}:
\begin{gather*}%\label{q252}
2\la_a\chi_{a,0}=T_{a,-1}=\langle
L^{-1}L^{0}\rangle-2P_a=-2(H_a+P_a),
\end{gather*} where $P_a$ is the density of the Hamiltonian of the shift
operator along $x$ corresponding to the  $a^{{\rm th}}$ site
$($marked point$)$:
\begin{gather}\label{q253}
\left\{\oint_{{\mathbb S}^1}\hbox{d}x\; P_a(x),
S^b(y)\right\}=k\delta_{ab}\p_y S^a,\qquad
P_a(x)=-\frac{k}{2}L_{11}^{a,-1}(x)\frac{\p_xL^{a,-1}_{12}(x)}{L^{a,-1}_{12}(x)}.
\end{gather}
\end{lemma}

\begin{proof}
The nontrivial part of the proof is related to the shift operators
$P_a$. Brackets (\ref{q3'x}) have the following form in standard
basis (here we omit index $a$ as above):
\begin{gather*}%\label{q255}
\{L^{-1}_{12}(x),L^{-1}_{11}(y)\}=2L^{-1}_{12}(x)\delta(x-y),\qquad
\{L^{-1}_{21}(x),L^{-1}_{11}(y)\}=-2L^{-1}_{21}(x)\delta(x-y),
\\
\{L^{-1}_{12}(x),L^{-1}_{21}(y)\}=-4L^{-1}_{11}(x)\delta(x-y).
\end{gather*}

For example, let us verify that
$\oint_{{\mathbb S}^1}\hbox{d}x\left\{L^{-1}_{11}(x)\frac{\p_xL^{-1}_{12}(x)}{L^{-1}_{12}(x)},L^{-1}_{21}(y)\right\}=-2\p_y L^{-1}_{21}(y)$:
\begin{gather*}%\label{q256}
\oint_{{\mathbb S}^1}\hbox{d}x
\left\{L^{-1}_{11}(x)\frac{\p_xL^{-1}_{12}(x)}{L^{-1}_{12}(x)},L^{-1}_{21}(y)\right\} \\
\qquad{}=\oint_{{\mathbb S}^1}\hbox{d}x\;
\big\{L^{-1}_{11}(x),L^{-1}_{21}(y)\big\}\frac{\p_xL^{-1}_{12}(x)}{L^{-1}_{12}(x)}-\oint_{{\mathbb
S}^1}\hbox{d}x\; \p_x L^{-1}_{11}(x)\{\ln L^{-1}_{12}(x),L^{-1}_{21}(y)\} \\
\qquad{}=2L^{-1}_{21}(y)\frac{\p_yL^{-1}_{12}(y)}{L^{-1}_{12}(y)}+4L^{-1}_{11}(y)
\frac{\p_yL^{-1}_{11}(y)}{L^{-1}_{12}(y)}=
-2\p_y L^{-1}_{21}(y).
 \end{gather*}

The later follows from the condition
$\p_y\la_a^2=0=\p_y((L^{-1}_{11}(y))^2+L^{-1}_{12}(y)L^{-1}_{21}(y))$.
The verif\/ication of (\ref{q253}) for other components can be
performed in the same way.
\end{proof}

\begin{lemma}\label{lemma4.2}
The density of the Hamiltonian $\chi_{a,1}$ has the following form
in terms of the decomposition of the $L$-matrix \eqref{q231}:
\begin{gather}
8\la_a^3\chi_{a,1}=4\la_a^2 T_{a,0}-T_{a,-1}^2 \nonumber \\
\phantom{8\la_a^3\chi_{a,1}} {}=2\langle L^{a,-1}L^{a,-1}\rangle\left(\frac{1}{2}\langle
L^{a,0}L^{a,0}\rangle+ \langle
L^{a,1}L^{a,-1}\rangle\right)-\left(\langle
L^{a,-1}L^{a,0}\rangle\right)^2 \nonumber\\
\phantom{8\la_a^3\chi_{a,1}=} {} +2k\langle L^{a,0}\p_x L^{a,-1}L^{a,-1}\rangle -\frac{k^2}{2}\langle
(\p_x L^{a,-1})^2\rangle.\label{q26}
\end{gather}
\end{lemma}

\begin{proof}
Our purpose is to show that $\oint_{{\mathbb S}^1}\hbox{d}x\;
\hbox{l.h.s.} \ (\ref{q26})=\oint_{{\mathbb S}^1}\hbox{d}x\;
\hbox{r.h.s.} \ (\ref{q26})$. The proof is direct. It is based on the
integration by parts. For example,
\begin{gather*}%\label{q27}
\oint_{{\mathbb S}^1}\hbox{d}x\;
4\la_a^2\left(-\frac{k^2}{2}\frac{\p_x^2
L_{12}^{-1}}{L_{12}^{-1}}+\frac{3k^2}{4}\left(\frac{\p_x
L_{12}^{-1}}{L_{12}^{-1}}\right)\right)-k^2\left(\frac{L_{11}^{-1}\p_x
L_{12}^{-1}}{L_{12}^{-1}}-\p_x
L_{11}^{-1}\right)^2 \\
\qquad{}=-\oint_{{\mathbb S}^1}\hbox{d}x\;
\frac{k^2}{2}\langle L_{x}^{-1}L_{x}^{-1}\rangle
\end{gather*}
and
\begin{gather*}%\label{q28}
\oint_{{\mathbb S}^1}\hbox{d}x\;
4\la_a^2\frac{k}{L_{12}^{-1}}\left(L_{11}^0\p_x
L_{12}^{-1}+L_{11}^{-1}\p_x L_{12}^0-\frac{L_{12}^0L_{11}^{-1}\p_x
L_{12}^{-1}}{L_{12}^{-1}}\right)
\\
{}-2k\left(\frac{L_{11}^{-1}\p_x L_{12}^{-1}}{L_{12}^{-1}}-\p_x
L_{11}^{-1}\right)(L_{12}^{-1}L_{21}^{0}+L_{12}^{0}L_{21}^{-1}+2L_{11}^{0}L_{11}^{-1})=
2k\oint_{{\mathbb S}^1}\hbox{d}x\; \langle
L^0L_x^{-1}L^{-1}\rangle.\!\!\!\!\tag*{\qed}
\end{gather*}
\renewcommand{\qed}{}
\end{proof}

Finally, we have the following densities of the Hamiltonians
describing the f\/irst and the second f\/lows:
\begin{gather}
h_{a,1}=-\la_a\chi_{a,0}=-\frac{1}{2}T_{a,-1}=P_a-\frac{1}{2}\langle
L^{a,-1}L^{a,0}\rangle,\nonumber
\\
h_{a,2}=-\la_a\chi_{a,1}=-\frac{1}{2}\left(T_{a,0}-\frac{1}{4\la_a^2}T_{a,-1}^2\right)=
-\frac{1}{4}\langle L^{a,0}L^{a,0}\rangle
\nonumber\\
{}-\frac{1}{2}\langle L^{a,1}L^{a,-1}\rangle+\frac{1}{8\la_a^2}\langle
L^{a,-1}L^{a,0}\rangle^2-\frac{k}{4\la_a^2}\langle L^{a,0}\p_x
L^{a,-1}L^{a,-1}\rangle +\frac{k^2}{16\la_a^2}\langle (\p_x
L^{a,-1})^2\rangle.\label{q29}
\end{gather}

These equalities are understood in a sense that
$\oint_{{\mathbb S}^1}\hbox{d}x\;
\hbox{l.h.s.} \ (\ref{q29})=\oint_{{\mathbb S}^1}\hbox{d}x\;
\hbox{r.h.s.} \ (\ref{q29})$.

\begin{theorem}\label{theorem4.1}
The Hamiltonian densities \eqref{q29} and the Poisson structure
\eqref{q3'x} provides equations \eqref{q15} and \eqref{q181}.
\end{theorem}

\begin{proof}
Substituting (\ref{q21}) into (\ref{q29}) we obtain explicit
expressions for the Hamiltonians. In rational case:
\begin{gather}\label{q30}
\mathcal{H}_{a,1}=\oint_{{\mathbb S}^1}\hbox{d}x\;
h_{a,1}(x)= \oint_{{\mathbb S}^1}\hbox{d}x\;
\left(P_a-\frac{1}{2}\sum_{c\neq a}\frac{\langle
S^aS^c\rangle}{z_a-z_c}\right)=\oint_{{\mathbb
S}^1}\hbox{d}x\; (P_a+H_a),
\\
%\label{q31}
\mathcal{H}_{a,2}=\oint_{{\mathbb
S}^1}\hbox{d}x\; h_{a,2}(x)\nonumber\\
\phantom{\mathcal{H}_{a,2}}{} =\oint_{{\mathbb S}^1}\hbox{d}x\;
\left(\frac{1}{2}\sum\limits_{c\neq a}\frac{\langle
S^aS^c\rangle}{(z_a-z_c)^2}-
\frac{1}{4}\langle\left(\sum\limits_{c\neq
a}\frac{S^c}{z_a-z_c}\right)^2\rangle+\frac{1}{8\la_a^2}\left(\sum\limits_{c\neq
a}\frac{\langle S^aS^c\rangle}{z_a-z_c}\right)^2\right.
\nonumber\\
\left. \phantom{\mathcal{H}_{a,2}=}{} - \frac{k}{4\la_a^2}\sum\limits_{c\neq a}\frac{\langle
S^c\p_xS^aS^a\rangle}{z_a-z_c}+\frac{k^2}{16\la_a^2}\langle\left(\p_xS^a\right)^2\rangle
\right)
\nonumber\\
\phantom{\mathcal{H}_{a,2}}{} =\oint_{{\mathbb S}^1}\hbox{d}x\; \left(\ti{H}_a-
\frac{k}{4\la_a^2}\sum\limits_{c\neq a}\frac{\langle
S^c\p_xS^aS^a\rangle}{z_a-z_c}+\frac{k^2}{16\la_a^2}\langle\left(\p_xS^a\right)^2\rangle
\right)\nonumber
\end{gather}
and in the elliptic case:
\begin{gather*}%\label{q32}
\mathcal{H}_{a,1}=\oint_{{\mathbb S}^1}\hbox{d}x\;
h_{a,1}(x)= \oint_{{\mathbb S}^1}\hbox{d}x\;
\left(P_a-\frac{1}{2}\sum\limits_{c\neq a}\langle
S^a\hat\vf_{ac}(S^c)\rangle\right)=\oint_{{\mathbb
S}^1}\hbox{d}x\; (P_a+H_a), \\
%\label{q33}
 \mathcal{H}_{a,2}=\oint_{{\mathbb
S}^1}\hbox{d}x\; h_{a,2}(x)=\oint_{{\mathbb S}^1}\hbox{d}x\;
\left(\frac{1}{4}\langle
S^a\hat\wp(S^a)\rangle+\frac{1}{2}\sum_{c\neq a}\langle S^a
\hat{F}(S^c)\rangle-\frac{1}{4}\langle\left(\sum_{c\neq
a}\hat\vf_{ac}(S^c)\right)^2\rangle\right.
\\
\left.\phantom{\mathcal{H}_{a,2}=}{} +\frac{1}{8\la_a^2}\left(\sum\limits_{c\neq a}\langle
S^a\hat\vf_{ac}(S^c)\rangle\right)^2-
\frac{k}{4\la_a^2}\sum\limits_{c\neq a}\langle
\hat\vf_{ac}(S^c)\p_xS^aS^a\rangle+\frac{k^2}{16\la_a^2}\langle\left(\p_xS^a\right)^2\rangle
\right)
\\
\phantom{\mathcal{H}_{a,2}}{} =\oint_{{\mathbb S}^1}\hbox{d}x\; \left(\ti{H}_a-
\frac{k}{4\la_a^2}\sum\limits_{c\neq a}\langle
\hat\vf_{ac}(S^c)\p_xS^aS^a\rangle+\frac{k^2}{16\la_a^2}\langle\left(\p_xS^a\right)^2\rangle
\right),
\end{gather*} where in $H_a$, $\ti{H}_a$ from
(\ref{q315}), (\ref{q316}) and
(\ref{q116}) $S^a=S^a(x)$ is assumed. The rest of the proof is
simple. One should write equations of motion generated by the
obtained Hamiltonians and the Poisson structure~(\ref{q3'x}) and
verify that they coincide with equations~(\ref{q15}) and
(\ref{q181}). This evaluation is direct.
\end{proof}

\appendix

\section{Elliptic functions}\label{appendixA}

\subsection{Basic def\/initions and properties}\label{appendixA1}

Most of definitions are borrowed from~\cite{We,Ma}. We assume that $q=\exp (2\pi i\tau)$, where $\tau$ is the modular
parameter of the elliptic curve $\Sigma_\tau$ which is realized as
${\mathbb C}/\Gamma_\tau$, $\Gamma_\tau=\mZ\oplus\mZ\tau$.

The basic element is the theta  function $(\bfe=\exp 2\pi\imath)$:
\begin{gather}
\vth(z|\tau)=q^{\frac {1}{8}}\sum_{n\in {\bf Z}}(-1)^n\bfe\left(\oh
n(n+1)\tau+nz\right) \nonumber\\
\phantom{\vth(z|\tau)}{} = q^{\frac{1}{8}}e^{-\frac{i\pi}{4}} (e^{i\pi z}-e^{-i\pi z})
\prod_{n=1}^\infty(1-q^n)(1-q^ne^{2i\pi z})(1-q^ne^{-2i\pi z}).\label{A.1a}
\end{gather}

{\it The  Eisenstein functions}
\begin{gather*}%\label{A.1}
E_1(z|\tau)=\p_z\log\vth(z|\tau),
\qquad E_1(z|\tau)\sim\f1{z}-2\eta_1z,
\end{gather*} where
\begin{gather*}%\label{A.6}
\eta_1(\tau)=\frac{3}{\pi^2}
\sum_{m=-\infty}^{\infty}\sum_{n=-\infty}^{\infty '}
\frac{1}{(m\tau+n)^2}=\frac{24}{2\pi
i}\frac{\eta'(\tau)}{\eta(\tau)}
\end{gather*} and $
\eta(\tau)=q^{\frac{1}{24}}\prod_{n>0}(1-q^n)  $ is the Dedekind
function
\begin{gather*}%\label{A.2}
E_2(z|\tau)=-\p_zE_1(z|\tau)= \p_z^2\log\vth(z|\tau),
\qquad E_2(z|\tau)\sim\f1{z^2}+2\eta_1.
\end{gather*}
 The higher
Eisenstein functions
\begin{gather*}%\label{A.2a}
E_j(z)=\frac{(-1)^j}{(j-1)!}\p^{(j-2)}E_2(z), \qquad j>2.
\end{gather*}

It is easy to see that the even-numbered functions are even and the
odd-numbered ones are odd:
\begin{gather}\label{A.2ax}
E_{2k}(-z)=E_{2k}(z),\qquad E_{2k+1}(-z)=-E_{2k+1}(z).
\end{gather}

{\it Relation to the Weierstrass functions}
\begin{gather}\label{ax100}
\zeta(z,\tau)=E_1(z,\tau)+2\eta_1(\tau)z, \\
\label{ax101} \wp(z,\tau)=E_2(z,\tau)-2\eta_1(\tau).
\end{gather}

The next important function is
\begin{gather}\label{A.3} \phi(u,z)=
\frac {\vth(u+z)\vth'(0)} {\vth(u)\vth(z)},\\
 \label{A.300}
\phi(u,z)=\phi(z,u) ,\qquad \phi(-u,-z)=-\phi(u,z) .
\end{gather}
It
has a pole at $z=0$ and \begin{gather*}%\label{A.3a}
\phi(u,z)=\frac{1}{z}+E_1(u)+\frac{z}{2}(E_1^2(u)-\wp(u))+\cdots.
\end{gather*}

Let $f(u,z)=\p_u\phi(u,z)$. Then
\begin{gather}\label{A3b}
f(u,z)=\phi(u,z) (E_1(u+z)-E_1(u)).
\end{gather}

{\it Heat equation}
\begin{gather*}%\label{A.4b}
\p_\tau\phi(u,w)-\f1{2\pi i}\p_u\p_w\phi(u,w)=0.
\end{gather*}

{\it Quasi-periodicity}
\begin{alignat}{3}%\label{A.11}
& \vth(z+1)=-\vth(z) , \qquad && \vth(z+\tau)=-q^{-\oh}e^{-2\pi
iz}\vth(z), & \nonumber\\
\label{A.12}
& E_1(z+1)=E_1(z) ,\qquad && E_1(z+\tau)=E_1(z)-2\pi i ,&
\\
 %\label{A.13}
& E_2(z+1)=E_2(z) ,\qquad && E_2(z+\tau)=E_2(z) , & \nonumber\\
 %\label{A.14}
& \phi(u,z+1)=\phi(u,z) ,\qquad && \phi(u,z+\tau)=e^{-2\pi \imath
u}\phi(u,z), & \nonumber\\
%\label{A.15}
& f(u,z+1)=f(u,z) ,\qquad && f(u,z+\tau)=e^{-2\pi \imath
u}f(u,z)-2\pi\imath\phi(u,z) . & \nonumber
\end{alignat}

{\it  The Fay three-section formula:}
\begin{gather}\label{ad3}
\phi(u_1,z_1)\phi(u_2,z_2)-\phi(u_1+u_2,z_1)\phi(u_2,z_2-z_1)-
\phi(u_1+u_2,z_2)\phi(u_1,z_1-z_2)=0.
\end{gather} Particular case
of this formula is the  Calogero functional equation
\begin{gather}\label{ad2}
\phi(u,z)\p_v\phi(v,z)-\phi(v,z)\p_u\phi(u,z)=(E_2(u)-E_2(v))\phi(u+v,z),
\\
\label{ad22} \phi(u,z)\phi(-u,z)=E_2(z)-E_2(u),\\
\label{ad23} \phi(z,u_1)\phi(z,u_2)=\phi(z,u_1+u_2)(E_1(z)+E_1(u_1)+E_1(u_2)-E_1(z+u_1+u_2)).
\end{gather}

Another important relation is
\begin{gather}
\phi(v,z-w)\phi(u_1-v,z)\phi(u_2+v,w)
-\phi(u_1-u_2-v,z-w)\phi(u_2+v,z)\phi(u_1-v,w) \nonumber\\
\qquad{}
=\phi(u_1,z)\phi(u_2,w)f(u_1,u_2,v),\label{ir}
 \end{gather} where
\begin{gather*}%\label{ir1}
f(u_1,u_2,v)=E_1(v)-E_1(u_1-u_2-v)+E_1(u_1-v)-E_1(u_2+v).
\end{gather*}
Taking limit $u_2\rightarrow 0$ in (\ref{ir}) we obtain:
\begin{gather*}%\label{ir21}
\phi(v,z-w)\phi(u_1-v,z)\phi(v,w)
-\phi(u_1-v,z-w)\phi(v,z)\phi(u_1-v,w)\\
\qquad{}=
\phi(u_1,z)(E_2(v)-E_2(u_1-v)),
\end{gather*} which is equivalent to
(\ref{ad2}) due to (\ref{A3b}).

{\it  Theta functions with characteristics.}
For $a, b \in \Bbb Q$ by def\/inition:
\begin{gather*}\theta{\left[\begin{array}{c}
a\\
b
\end{array}
\right]}(z , \tau ) =\sum_{j\in \Bbb Z} {\bf
e}\left((j+a)^2\frac\tau2+(j+a)(z+b)\right).
\end{gather*} In
particular, the function $\vth$ (\ref{A.1a}) is a theta function
with characteristics:
\begin{gather*}%\label{A.29}
\vartheta(x,\tau)=\theta\left[
\begin{array}{c}
1/2\\
1/2
\end{array}\right](x,\tau).
\end{gather*} Properties:
\begin{gather*}
\theta{\left[\begin{array}{c}
a\\
b
\end{array}
\right]}(z+1 , \tau )={\bf e}(a) \theta{\left[\begin{array}{c}
a\\
b
\end{array}
\right]}(z  , \tau ) , \\
\theta{\left[\begin{array}{c}
a\\
b
\end{array}
\right]}(z+a'\tau , \tau ) ={\bf e}\left(-{a'}^2\frac\tau2
-a'(z+b)\right) \theta{\left[\begin{array}{c}
a+a'\\
b
\end{array}
\right]}(z , \tau ), \\
\theta{\left[\begin{array}{c}
a+j\\
b
\end{array}
\right]}(z , \tau )= \theta{\left[\begin{array}{c}
a\\
b
\end{array}
\right]}(z , \tau ),\qquad j\in \Bbb Z.
 \end{gather*}
 The following
notations are used:
$\theta\left[\begin{array}{l}a/2\\b/2\end{array}\right](z)=\theta_{ab}(z)$.
Then $\theta_{11}(z)=\vth(z)$ from (\ref{A.1a}).

\section[Lie algebra $\sln$ and elliptic functions]{Lie algebra $\boldsymbol{\sln}$ and elliptic functions}\label{appendixB}

Introduce the notation
\begin{gather*}
{\bf e}_N(z)=\exp \left(\frac{2\pi i}{N} z\right)
 \end{gather*}
 and two matrices
\begin{gather}\label{q} Q=\di({\bf e}_N(1),\ldots,{\bf e}_N(m),\ldots,1),\\
 \label{la} \La= \left(\begin{array}{ccccc}
0&1&0&\cdots&0\\
0&0&1&\cdots&0\\
\vdots&\vdots&\ddots&\ddots&\vdots\\
0&0&0&\cdots&1\\
1&0&0&\cdots&0
\end{array}\right) .
\end{gather} Let
\begin{gather}\label{B.10}
\Ga_N=\mZ^{(2)}_N=(\mZ/N\mZ\oplus\mZ/N\mZ) ,\qquad \Ga_N'=\ti{\mZ}^{(2)}_N=
\mZ^{(2)}_N\setminus(0,0)
\end{gather} be the two-dimensional
lattices of orders $N^2$ and $N^2-1$ correspondingly. The generators
of the lattice $\Ga_N$ corresponding to the elliptic curve
$\Sigma_\tau$ (generated by the lattice $\Gamma_\tau$) are
$\frac{1}{N}$ and $\frac{1}{\tau}$. The matrices $Q^{a_1}\La^{a_2}$,
$a=(a_1,a_2)\in\mZ^{(2)}_N$ generate a basis in the group $\GLN$,
while $Q^{\al_1}\La^{\al_2}$, $\al=(\al_1,\al_2)\in\ti{\mZ}^{(2)}_N$
generate a basis in the Lie algebra $\sln$. Consider the projective
representation of $\mZ^{(2)}_N$ in $\GLN$
\begin{gather}\label{B.11}
a\to T_{a}= \bfe_N\left(\frac{a_1a_2}{2}\right)Q^{a_1}\La^{a_2}, \\
\label{AA3a} T_aT_b=\bfe_N\left(-\frac{a\times b}{2}\right)T_{a+b}, \qquad (a\times
b=a_1b_2-a_2b_1).
\end{gather}
The natural Killing form is
\begin{gather}\label{Kil}
\langle T_\al
T_\be\rangle\, \overset{\rm def}{=}\, \hbox{Tr}(T_\al
T_\be)=N\delta_{0,\al+\be\ \hbox{mod}\ \Ga_N}.
 \end{gather} It
follows from (\ref{AA3a}) that
\begin{gather}\label{AA3b}
[T_{\al},T_{\be}]=c_{\al,\be}T_{\al+\be} ,
\end{gather}
 where
\begin{gather*}%\label{q1001}
c_{\al,\be}=2\sqrt{-1}\sin\frac{\pi}{N}(\al\times \be)
\end{gather*}
are the structure constants of $\sln$. They obey the following
properties: \begin{gather*}%\label{q1002}
c_{\al,\be}=c_{\al,\be+\al},\qquad
c_{\al,\be}=-c_{\be,\al}=c_{\be,-\al} .
\end{gather*}

Introduce the following  constants on $\ti{\mZ}^{(2)}$:
\begin{gather}%\label{AA50}
\vth(\om_\ga)=\vth\left(\frac{\ga_1+\ga_2\tau}{N}\right) ,
\nonumber\\
\label{AA5}
E_1(\om_\ga)=E_1\left(\frac{\ga_1+\ga_2\tau}{N}\right) ,
\qquad\!\! E_2(\om_\ga)=E_2\left(\frac{\ga_1+\ga_2\tau}{N}\right) , \qquad \!\!
 \wp(\om_\ga)=\wp\left(\frac{\ga_1+\ga_2\tau}{N}\right) \!\!\!\!\!
\end{gather}
and the quasi-periodic functions on $\Si_\tau$
\begin{gather}\label{phi}
\phi_\ga(z)=\phi(\frac{\ga_1+\ga_2\tau}{N},z) ,
\\
\label{vf} \vf_\ga(z)=\bfe_N(\ga_2z)\phi_\ga(z) ,
\\
\label{f} f_\ga(z)=
\bfe_N(\ga_2z)\p_u\phi(u,z)|_{u=\frac{\ga_1+\ga_2\tau}{N}}
=\vf_\ga(z)(E_1(\om_\ga+z)- E_1(\om_\ga))
, \\
\label{FF} F_\ga(z)=\vf_\ga(z)E_1(z)-f_\ga(z)= \vf_\ga(z)(E_1(z)+
E_1(\om_\ga)-E_1(\om_\ga+z)).
\end{gather} Function $\vf_\ga(z)$ is
an element of the basis in the space of sections with a simple pole
at $z=0$ of the bundle End$(V)$ for the holomorphic vector bundle
$V$ of degree~1. It follows from (\ref{A.3}) that
\begin{gather}\label{qp} \vf_\ga(z+1)=\bfe_N(\ga_2)\vf_\ga(z),\qquad
\vf_\ga(z+\tau)=\bfe_N(-\ga_1)\vf_\ga(z),
\\
\label{qpf} f_\ga(z+1)=\bfe_N(\ga_2)f_\ga(z) ,\qquad
f_\ga(z+\tau)=\bfe_N(-\ga_1)f_\ga(z)-2\pi\imath\vf_\ga(z),\\
\label{qpF} F_\ga(z+1)=\bfe_N(\ga_2)F_\ga(z) ,\qquad
F_\ga(z+\tau)=\bfe_N(-\ga_1)F_\ga(z). \end{gather}
Function
$F_\ga(z)$ is the quasi-periodic and has the second order pole at
$z=0$.

Let us write down the Fay-type formulae. It follows from (\ref{ad3})--(\ref{ad23}) that
\begin{gather}\label{we4}
\vf_\ga(z-z_a)\vf_\be(z-z_c)\; \overset{\eqref{ad3}}{=}\;
\vf_{\be+\ga}(z-z_a)\vf_\be(z_a-z_c)+\vf_{\be+\ga}(z-z_c)\vf_\ga(z_c-z_a),
\\
\label{we5}
\vf_\be(z)f_\ga(z)-\vf_\ga(z)f_\be(z)\; \overset{\eqref{ad2}}{=}\;
\vf_{\be+\ga}(z)(\wp(\om_\be)-\wp(\om_\ga)),
\\
\label{we6}
\vf_\al(z)\vf_{-\al}(z) \; \overset{\eqref{ad22}}{=}\; \wp(z)-\wp(\om_\al),
\\
\vf_\be(z)\vf_\ga(z)\; \overset{\eqref{ad23}}{=}\;
\vf_{\be+\ga}(z)(E_1(z)+E_1(\om_\be)+E_1(\om_\ga)-E_1(z+\om_\be+\om_\ga)),
\nonumber\\
\vf_\ga(z_1)\vf_\ga(z_2)\; \overset{\eqref{ad23}}{=}\;
\vf_\ga(z_1+z_2)(E_1(z_1)+E_1(z_2)+E_1(\om_\ga)-E_1(z_1+z_2+\om_\ga)).\label{we7}
\end{gather}

The last one identity can be also rewritten (using  (\ref{A.300})
which is $\vf_\ga(-z)=-\vf_{-\ga}(z)$ and (\ref{A.2ax})) as follows:
\begin{gather*}
\vf_\ga(z_1)\vf_{-\ga}(z_2)=
-\vf_\ga(z_1-z_2)(E_1(z_1)-E_1(z_2)+E_1(\om_\ga)-E_1(z_1-z_2+\om_\ga)).
\end{gather*}

We also need the following relation:
\begin{gather}
-\vf_\be(z-z_c)f_\ga(z-z_a)+\vf_\ga(z-z_a)f_\be(z-z_c)\nonumber
\\
\qquad{}=-\vf_{\be+\ga}(z-z_c)f_\ga(z_c-z_a)+\vf_{\be+\ga}(z-z_a)f_\be(z_a-z_c).\label{we8}
\end{gather}

\section[Lie algebra $\sl2$ and elliptic functions]{Lie algebra $\boldsymbol{\sl2}$ and elliptic functions}\label{appendixC}

 For   $\SLT$  instead of $T_\al$ we use the basis
of sigma-matrices \begin{gather*}%\label{100}
 \si_0={\rm Id}, \qquad \si_1=
T_{0,1}, \qquad \si_2= T_{1,1}, \qquad \si_3=- T_{1,0}, \\
\{\si_a\}=\{\si_0,\si_\al\}, \qquad a=0,\al, \qquad \al=1,2,3,
\\
\si_+=\frac{\si_1-\imath\si_2}2, \qquad \si_-=\frac{\si_1+\imath\si_2}2.
\end{gather*}
The standard theta-functions with the characteristics
are
\begin{gather*}%\label{101}
\te_{0,0}=\te_3,\qquad \te_{1,0}= \te_2, \qquad \te_{0,1}=\te_4, \qquad \te_{1,1}=\te_1.
\end{gather*}

For $\al=1,2,3$ and
$\{\om_\al\}=\{\frac{\tau}{2},\frac{\tau+1}{2},\frac{1}{2}\}$
\begin{gather*}%\label{a1}
\vf_\al(z)=\bfe(z\p_\tau\om_\al)\frac{\vth'(0)\vth(z+\om_\al)}{\vth(z)\vth(\om_\al)}.
\end{gather*}

\begin{table}[th!]
\centering
\caption{}
\vspace{1mm}

\begin{tabular}{|c||c|c|c|}
\hline
$\al$ & (1,0) & (0,1) & (1,1) \\
\hline $\si_\al$ & $\si_3$   & $\si_1$ &
$\si_2$ \\
\hline
\tsep{2pt} half-periods & $\om_1=\oh$ & $\om_2=\frac{\tau}{2}$ &
$\om_3=\frac{1+\tau}{2}$ \bsep{2pt} \\
\hline
\tsep{4pt} $\varphi_\al (z)$    &
$\frac{\te_2(z)\te'_1(0)}{\te_2(0)\te_1(z)}$ &
$\frac{\te_4(z)\te'_1(0)}{\te_4(0)\te_1(z)}$ &
$\frac{\te_3(z)\te'_1(0)}{\te_3(0)\te_1(z)} $\bsep{4pt}\\
\hline
\end{tabular}
\end{table}

In $\sl2$ case some more properties appear in addition to the
previously listed. In what follows $\al$, $\be$, $\ga$ are dif\/ferent
indices equivalent to $1$, $2$, $3$ up to a cyclic permutation. Then
$\om_\al+\om_\be=\om_\ga$ $\mod \Gamma_2$ and
$\vf_{\al+\be}(z)=\vf_\ga(z)$,
\begin{gather}\label{a2}
 \vf_{-\al}(z)=\vf_\al(z),\qquad
\vf_\al(-z)=-\vf_\al(z),
 \\
\label{a21} E_1(\om_\al)=-2\pi\sqrt{-1}\p_\tau\om_\al.
 \end{gather}
 Indeed, from
(\ref{A.2ax}) and (\ref{A.12}) we have
$-E_1(\frac{\tau}{2})=E_1(\frac{\tau}{2}-\tau)=E_1(\frac{\tau}{2})+2\pi\sqrt{-1}$.
Then
\begin{gather}\label{a22}
E_1(\om_\al)+E_1(\om_\be)=E_1(\om_\al+\om_\be) ,\\
\label{a3} (\vf_\al(z))^2=\wp(z)-\wp(\om_\al).
\end{gather}

For small $z$:
\begin{gather*}\vf_\al(z)=\frac{1}{z}-\frac{z}{2}\wp(\om_\al)+\cdots.
\end{gather*}

From (\ref{a22}) we also have
\begin{gather}\label{a31}
F_\al(z)=\vf_\be(z)\vf_\ga(z)=-\p_z\vf_\al(z).
\end{gather}

The Fay identity (\ref{ad3}) reads:
\begin{gather}\label{a4}
\vf_\ga(z-z_a)\vf_\be(z-z_c)=\vf_\al(z-z_a)\vf_\be(z_a-z_c)-\vf_\al(z-z_c)\vf_\ga(z_a-z_c).
\end{gather}

Combining (\ref{a4}) we may get:
\begin{gather}
\vf_\be(z-z_c)\vf_\be(z-z_a)\vf_\al(z-z_a)=\vf_\be(z-z_a)\vf_\ga(z-z_a)\vf_\be(z_a-z_c)\nonumber
\\
\qquad{}
+\vf_\al(z-z_c)\vf_\al(z_c-z_a)\vf_\be(z_c-z_a)-\vf_\al(z-z_a)\vf_\al(z_c-z_a)\vf_\ga(z_c-z_a)\label{a5}
\end{gather} or
\begin{gather}
\vf_\ga(z-z_c)\vf_\al(z-z_a)\vf_\ga(z-z_a)=\vf_\be(z-z_a)\vf_\ga(z-z_a)\vf_\ga(z_a-z_c)
\nonumber\\
\qquad{} +\vf_\al(z-z_a)\vf_\al(z_c-z_a)\vf_\be(z_a-z_c)-\vf_\al(z-z_c)\vf_\al(z_c-z_a)\vf_\ga(z_a-z_c).\label{a6}
\end{gather}

\subsection*{Acknowledgements}

Author is grateful to M.A.~Olshanetsky for useful discussions and
remarks. The work was supported by grants RFBR-09-02-00393,
RFBR-09-01-92437-KEa, RFBR-09-01-93106-NCNILa, Russian President
fund MK-1646.2011.1 and to the Federal Agency for Science and
Innovations of Russian Federation under contract 14.740.11.0347.

\pdfbookmark[1]{References}{ref}
\LastPageEnding


\begin{thebibliography}{99}

\footnotesize\itemsep=0pt

\bibitem{Gaudin1}
%Gaudin M.,
%Mod{\`e}les exacts en m{\`e}canique statistique: la m{\`e}thode de Bethe et ses g{\`e}n{\`e}ralizations,
%{\it Note CEA} {\bf 1559} (1972) (1), {\it Note CEA} {\bf 1559} (1973) (2).\\
 Gaudin M.,
 Diagonalisation d{\`u}ne classe d'Hamiltoniens de spin,
\href{http://dx.doi.org/10.1051/jphys:0197600370100108700}{{\it J.~Physique}} {\bf  37} (1976), 1087--1098.\\
Gaudin M.,
La fonction d'onde de Bethe, Masson, Paris, 1983 (in French); Mir, Moscow, 1987 (in Russian).

\bibitem{Baxter1}
Baxter R.J.,
One-dimensional anisotropic Heisenberg chain,
\href{http://dx.doi.org/10.1016/0003-4916(72)90270-9}{{\it Ann. Physics}} {\bf  70} (1972), 323--337.

\bibitem{SklTak}
Sklyanin E.K., Takebe T.,
Algebraic bethe ansatz for the XYZ Gaudin model,
\href{http://dx.doi.org/10.1016/0375-9601(96)00448-3}{{\it Phys. Lett.~A}} {\bf 219} (1996), 217--225,
\href{http://arxiv.org/abs/q-alg/9601028}{q-alg/9601028}.

\bibitem{Zotov1}
Zotov  A.,
 Elliptic linear problem for Calogero--Inozemtsev model and Painlev{\'e} VI equation,
\href{http://dx.doi.org/10.1023/B:MATH.0000032753.97756.94}{{\it Lett. Math. Phys.}} {\bf 67} (2004), 153--165,
 \href{http://arxiv.org/abs/hep-th/0310260}{hep-th/0310260}.
\\
Levin A., Olshanetsky M., Zotov A.,
Painlev{\'e} VI, rigid tops and ref\/lection equation,
\href{http://dx.doi.org/10.1007/s00220-006-0089-y}{{\it Comm. Math. Phys.}} {\bf 268} (2006), 67--103,
\href{http://arxiv.org/abs/math.QA/0508058}{math.QA/0508058}.


\bibitem{LZ}
 Levin A., Zotov A.,
 On rational and elliptic forms of Painlev{\'e} VI equation,
{\it Amer. Math. Soc. Transl. Ser.~2}, Vol.~221, Amer. Math. Soc., Providence, RI, 173--184.

\bibitem{Schlez}
Fuchs R.,
\"Uber lineare homogene Dif\/ferentialgleichungen zweiterordnung mit im endlich gelegne wesentlich singul\"aren Stellen,
\href{http://dx.doi.org/10.1007/BF01449199}{{\it Math. Ann.}} {\bf 63} (1907), 301--323.
\\
Schlesinger L.,
\"Uber eine Klasse von Dif\/ferentialsystemen beliebiger Ordnung mit festen kritischen Punkten,
\href{http://dx.doi.org/10.1515/crll.1912.141.96}{{\it J. Reine Angew. Math.}} {\bf 141} (1912), 96--145.

\bibitem{STSR}
Reyman A.G., Semenov-Tian-Shansky M.A.,
Lie algebras and Lax equations with spectral parameter on an elliptic curve,
{\it Zap. Nauchn. Semin. Leningr. Otd. Mat. Inst. Steklova} {\bf 150} (1986), 104--118 (in Russian).

\bibitem{Belavin}
 Belavin A.A., Drinfel'd V.G.,
 Solutions of the classical Yang--Baxter equation for simple Lie algebras,
 {\it Funktsional. Anal. i Prilozhen.} {\bf 16} (1982), no.~3, 1--29
 (English transl.: \href{http://dx.doi.org/10.1007/BF01081585}{{\it Funct. Anal. Appl.}} {\bf 16} (1982), 159--180).

\bibitem{LOZCh}
Zotov A.V., Levin A.M.; Olshanetsky M.A., Chernyakov Yu.B.,
Quadratic algebras related to elliptic curves,
\href{http://dx.doi.org/10.1007/s11232-008-0081-0}{{\it Theoret. and Math. Phys.}} {\bf 156} (2008), 1103--1122,
\href{http://arxiv.org/abs/0710.1072}{arXiv:0710.1072}.

\bibitem{Skl}
Sklyanin E.K.,
Separation of variables in the Gaudin model,
{\it Zap. Nauchn. Sem. Leningrad. Otdel. Mat. Inst. Steklov. (LOMI)} {\bf 164} (1987), 151--169
(English transl.: {\it  J.~Soviet Math.} {\bf 47} (1989), 2473--2488).
\\
 Sklyanin E.K., Takebe T.,
 Separation of variables in the elliptic Gaudin model,
\href{http://dx.doi.org/10.1007/s002200050635}{{\it Comm. Math. Phys.}} {\bf 204} (1999), 17--38,
\href{http://arxiv.org/abs/solv-int/9807008}{solv-int/9807008}.

\bibitem{Skl2}
 Sklyanin E.K.,
 Generating function of correlators in the ${\rm sl}_2$ Gaudin model,
\href{http://dx.doi.org/10.1023/A:1007585716273}{{\it Lett. Math. Phys.}} {\bf 47} (1999), 275--292,
\href{http://arxiv.org/abs/solv-int/9708007}{solv-int/9708007}.
\\
Takasaki K.,
Gaudin model, KZ equation and an isomonodromic problem on the tours,
\href{http://dx.doi.org/10.1023/A:1007417518021}{{\it Lett. Math. Phys.}} {\bf 44} (1998),  143--156,
\href{http://arxiv.org/abs/hep-th/9711058}{hep-th/9711058}.
\\
Feigin B., Frenkel E., Reshetikhin N.,
Gaudin model, Bethe ansatz and critical level,
\href{http://dx.doi.org/10.1007/BF02099300}{{\it Comm. Math. Phys.}} {\bf 166} (1994), 27--62,
\href{http://arxiv.org/abs/hep-th/9402022}{hep-th/9402022}.
\\
 Gould M.D., Zhang Y.-Z., Zhao S.-Y.,
 Elliptic Gaudin models and elliptic KZ equations,
\href{http://dx.doi.org/10.1016/S0550-3213(02)00136-0}{{\it Nuclear Phys.~B}} {\bf 630} (2002), 492--508,
\href{http://arxiv.org/abs/nlin.SI/0110038}{nlin.SI/0110038}.
\\
Chernyakov Yu., Levin A., Olshanetsky M., Zotov A.,
Elliptic Schlesinger system and Painlev\'e VI,
\href{http://dx.doi.org/10.1088/0305-4470/39/39/S05}{{\it J.~Phys.~A: Math. Gen.}} {\bf 39} (2006), 12083--12101,
\href{http://arxiv.org/abs/nlin.SI/0602043}{nlin.SI/0602043}.

\bibitem{KLO}
Khesin B., Levin A., Olshanetsky M.,
Bihamiltonian structures and quadratic algebras in hydrodynamics and on non-commutative torus,
\href{http://dx.doi.org/10.1007/s00220-004-1150-3}{{\it Comm. Math. Phys.}} {\bf 250} (2004), 581--612,
\href{http://arxiv.org/abs/nlin.SI/0309017}{nlin.SI/0309017}.

\bibitem{discr}
Petrera M., Suris Yu.B.,
An integrable discretization of the rational ${\mathfrak{su}}(2)$ Gaudin model and related systems,
\href{http://dx.doi.org/10.1007/s00220-008-0512-7}{{\it Comm. Math. Phys.}} {\bf 283} (2008), 227--253,
\href{http://arxiv.org/abs/0707.4088}{arXiv:0707.4088}.\\
Ragnisco O., Zullo F.,
B\"acklund transformations for the trigonometric Gaudin magnet,
\href{http://dx.doi.org/10.3842/SIGMA.2010.012}{{\it SIGMA}} {\bf 6} (2010), 012, 6~pages,
\href{http://arxiv.org/abs/0912.2456}{arXiv:0912.2456}.
\\
Veselov A.P.,
What is an integrable mapping?,
in What is Integrability?, Editor V.E.~Zakharov,
{\it Springer Ser. Nonlinear Dynam.}, Springer, Berlin, 1991, 251--272.
\\
 Hone A.N.W., Kuznetsov V.B., Ragnisco O.,
 B\"acklund transformations for the ${\rm sl}(2)$ Gaudin magnet,
{\it J.~Phys.~A: Math. Gen.} {\bf 34} (2001), 2477--2490,
\href{http://arxiv.org/abs/nlin.SI/0007041}{nlin.SI/0007041}.

\bibitem{RST}
 Talalaev D.V.,
The quantum Gaudin system, {\it Funktsional. Anal. i Prilozhen.} {\bf 40} (2006), no.~1, 86--91 (Eng\-lish transl.:
\href{http://dx.doi.org/10.1007/s10688-006-0012-5}{{\it Funct. Anal. Appl.}} {\bf 40} (2006), 73--77).
\\
Rubtsov V., Silantyev A., Talalaev D.,
Manin matrices, quantum elliptic commutative families and characteristic polynomial of elliptic Gaudin model,
\href{http://dx.doi.org/10.3842/SIGMA.2009.110}{{\it SIGMA}} {\bf 5} (2009), 110, 22~pages,
\href{http://arxiv.org/abs/0908.4064}{arXiv:0908.4064}.

\bibitem{Langl}
Frenkel E.,
Af\/f\/ine algebras, Langlands duality and Bethe ansatz,
\href{http://arxiv.org/abs/q-alg/9506003}{q-alg/9506003}.
\\
Chervov A., Talalaev D.,
Quantum spectral curves, quantum integrable systems and the geometric Langlands correspondence,
\href{http://arxiv.org/abs/hep-th/0604128}{hep-th/0604128}.
\\
Teschner J.,
Quantization of the Hitchin moduli spaces, Liouville theory, and the geometric Langlands correspondence~I,
\href{http://arxiv.org/abs/1005.2846}{arXiv:1005.2846}.

\bibitem{ER}
 Enriquez  B., Rubtsov V.,
 Hitchin systems, higher Gaudin operators and $R$-matrices,
 {\it Math. Res. Lett.} {\bf  3} (1996), 343--357,
\href{http://arxiv.org/abs/alg-geom/9503010}{alg-geom/9503010}.


\bibitem{H}
Hitchin N.,
Stable bundles and integrable systems,
\href{http://dx.doi.org/10.1215/S0012-7094-87-05408-1}{{\it Duke Math.~J.}} {\bf 54} (1987), 91--114.

\bibitem{GN}
Nekrasov  N.,
Holomorphic bundles and many-body systems,
\href{http://dx.doi.org/10.1007/BF02099624}{{\it Comm. Math. Phys.}} {\bf 180} (1996), 587--603,
\href{http://arxiv.org/abs/hep-th/9503157}{hep-th/9503157}.
\\
Gorsky  A., Nekrasov N.,
 Elliptic Calogero--Moser system from two dimensional current algebra,
\href{http://arxiv.org/abs/hep-th/9401021}{hep-th/9401021}.

\bibitem{LOZ}
Levin A., Olshanetsky M., Zotov A.,
Hitchin systems -- symplectic Hecke correspondence and two-dimensional version,
\href{http://dx.doi.org/10.1007/s00220-003-0801-0}{{\it Comm. Math. Phys.}} {\bf 236} (2003), 93--133,
\href{http://arxiv.org/abs/nlin.SI/0110045}{nlin.SI/0110045}.


\bibitem{Ragnisco}
 Petrera M., Ragnisco O.,
 From $su(2)$ gaudin models to integrable tops,
 \href{http://dx.doi.org/10.3842/SIGMA.2007.058}{{\it SIGMA}} {\bf 3} (2007), 058, 14~pages,
\href{http://arxiv.org/abs/math-ph/0703044}{math-ph/0703044}.\\
Levin A., Zotov A.,
An integrable system of interacting elliptic tops,
{\it Teoret. Mat. Fiz.} {\bf 146} (2006), 55--64 (English transl.: \href{http://dx.doi.org/10.1007/s11232-006-0005-9}{{\it Theoret. and Math. Phys.}} {\bf  146} (2006), 45--52).

\bibitem{Kuznetsov}
 Kuznetsov V.B.,
 Isomorphism of the $n$-dimensional Neumann system and the $n$-site Gaudin magnet,
{\it Funktsional. Anal. i Prilozhen.} {\bf 26} (1992), no.~4, 88--90 (English transl.:
\href{http://dx.doi.org/10.1007/BF01075058}{{\it Funct. Anal. Appl.}} {\bf  26} (1992), 302--304).





\bibitem{ZSh}
Zakharov V.E., Shabat A.B.,
A scheme for integrating the nonlinear equations of mathematical physics by the method of the inverse scattering problem.~I,
{\it Funktsional. Anal. i Prilozhen.} {\bf 8} (1974), no.~3, 43--53 (English transl.:
\href{http://dx.doi.org/10.1007/BF01075696}{{\it Funct. Anal. Appl.}} {\bf  8} (1974), 226--235).

\bibitem{FT2}
Faddeev L., Takhtajan L.,
Hamiltonian approach to solitons theory, Nauka, Moscow, 1986 (in Russian).



\bibitem{DKN}
Dubrovin B.A., Krichever I.M., Novikov S.P.,
Integrable systems. I, {\it  Current Problems in Mathematics. Fundamental Directions}, Vol.~4,
Itogi Nauki i Tekhniki, Akad. Nauk SSSR, Vsesoyuz. Inst. Nauchn. i Tekhn. Inform., Moscow, 1985, 179--284 (in Russian).

\bibitem{MS}
Sokolov V.V., Schabat A.B.,
Classif\/ication of integrable evolution equations,
{\it Soviet Sci. Rev. Sect.~C Math. Phys. Rev.}, Vol.~4, Harwood Academic Publ., Chur, 1984, 221--280.\\
Mikhailov A.V., Schabat A.B., Yamilov R.I.,
The symmetry approach to classif\/ication of nonlinear equations. Complete list of integrable systems,
{\it Uspekhi Mat. Nauk} {\bf 42} (1987), no.~4, 3--53
(English transl.: \href{http://dx.doi.org/10.1070/RM1987v042n04ABEH001441}{{\it Russian Math. Surveys}} {\bf 42} (1987), no.~4, 1--63).\\
Fokas A.S., Symmetries and integrability,
{\it Stud. Appl. Math.} {\bf 77} (1987), 253--299.




\bibitem{LL}
Landau L., Lifshitz E.,
On the theory of the dispersion of magnetic permeability in ferromagnetic bodies,
{\it Phys. Zeitsch. der Sow.} {\bf 8} (1935), 153--169.


\bibitem{Sk}
Sklyanin E.,
On complete integrability of the Landau--Lifshitz equation,
Preprint LOMI E-3-79, 1979. \\
Borovik  A.E., Robuk V.N.,
Linear pseudopotentials and conservation laws for the Landau-Lifshits equation describing the nonlinear dynamics of a ferromagnet with uniaxial anisotropy,
{\it Teoret. Mat. Fiz.}  {\bf 46} (1981), 371--381 (English transl.: \href{http://dx.doi.org/10.1007/BF01032734}{{\it Theoret. and Math. Phys.}} {\bf 46} (1981), 242--248).

\bibitem{FT}
Baxter R.J.,
Eight-vertex model in lattice statistics and one-dimensional anisotropic Heisenberg chain. I.~Some fundamental eigenvectors,
\href{http://dx.doi.org/10.1016/0003-4916(73)90439-9}{{\it Ann. Physics}} {\bf 76} (1973), 1--24.\\
 Takhtajan L.A., Faddeev L.D.,
The quantum method of the inverse problem and the Heisenberg XYZ model,
{\it Uspekhi Mat. Nauk} {\bf 34} (1979), no.~5, 13--63
(English transl.: \href{http://dx.doi.org/10.1070/RM1979v034n05ABEH003909}{{\it Russian Math. Surveys}} {\bf 34} (1979), no.~5, 11--68).



\bibitem{ZakhMikh}
Zakharov V.E., Mikhailov A.V.,
Relativistically invariant two-dimensional models of f\/ield theory which are integrable by means of the inverse scattering problem method,
{\it Soviet Phys. JETP} {\bf 74} (1978), 1953--1973.
\\
 Pohlmeyer K.,
 Integrable Hamiltonian systems and interactions through quadratic constraints,
\href{http://dx.doi.org/10.1007/BF01609119}{{\it Comm. Math. Phys.}} {\bf 46} (1976), 207--221.

\bibitem{Chered}
Cherednik I.V.,
Local conservation laws of principal chiral f\/ields $(d=1)$,
{\it Teoret. Mat. Fiz.} {\bf 38} (1979), 179--185
(English transl.: \href{http://dx.doi.org/10.1007/BF01016832}{{\it Theoret. and Math. Phys.}} {\bf 38} (1979), 120--124).
\\
Cherednik I.V.,
Relativistically invariant quasiclassical limits of integrable two-dimensional quantum models,
{\it Teoret. Mat. Fiz.} {\bf 47} (1981), 225--229
(English transl.: \href{http://dx.doi.org/10.1007/BF01086395}{{\it Theoret. and Math. Phys.}} {\bf 47} (1981), 422--425).

\bibitem{K}
Krichever I.,
Vector bundles and Lax equations on algebraic curves,
\href{http://dx.doi.org/10.1007/s002200200659}{{\it Comm. Math. Phys.}} {\bf 229} (2002), 229--269,
\href{http://arxiv.org/abs/hep-th/0108110}{hep-th/0108110}.

\bibitem{DMN}
Dubrovin B.A., Matveev V.B., Novikov S.P.,
Non-linear equations of Korteweg--de Vries type, f\/inite-zone linear operators, and Abelian varieties,
{\it Uspekhi Mat. Nauk} {\bf 31} (1976), no.~1, 55--136
(English transl.: \href{http://dx.doi.org/10.1070/RM1976v031n01ABEH001446}{{\it Russian Math. Surveys}} {\bf 31} (1976), no.~1, 59--146).

\bibitem{GolSok}
Golubchik I.Z., Sokolov V.V.,
Multicomponent generalization of the hierarchy of the Landau--Lifshitz equation,
{\it Teoret. Mat. Fiz.} {\bf 124} (2000), 62--71
(English transl.: \href{http://dx.doi.org/10.1007/BF02551067}{{\it Theoret. and Math. Phys.}} {\bf 124} (2000), 909--917).



\bibitem{KrichVol}
Akhmetshin A.A.,  Krichever I.M., Volvovski Yu.S.,
Elliptic families of solutions of the Kadomtsev--Petviashvili equation, and the f\/ield analogue of the elliptic Calogero--Moser system,
{\it Funktsional. Anal. i Prilozhen.} {\bf 36} (2002), no.~4, 1--17 (English transl.:
\href{http://dx.doi.org/10.1023/A:1021706525301}{{\it Funct. Anal. Appl.}} {\bf 36} (2002),  253--266),
\href{http://arxiv.org/abs/hep-th/0203192}{hep-th/0203192}.

\bibitem{Skryp}
Skrypnik T.,
`Doubled' generalized Landau--Lifshiz hierarchies and special quasigraded Lie algebras,
\href{http://dx.doi.org/10.1088/0305-4470/37/31/008}{{\it J.~Phys.~A: Math. Gen.}} {\bf 37} (2004), 7755--7768.\\
Skrypnik T.,
Quasigraded Lie algebras and matrix generalization of Landau--Lifshitz equation,
in Proceedinds of Fifth International Conference ``Symmetry in
Nonlinear Mathematical Physics'' (June 23--29, 2003, Kyiv),
Editors A.G.~Nikitin, V.M.~Boyko, R.O.~Popovych and I.A.~Yehorchenko,
{\it Proceedings of Institute of Mathematics, Kyiv\/} {\bf 50} (2004), Part~1, 462--469.

\bibitem{Holod}
Holod P.I.,
The hidden symmetry of the Landau--Lifshits equation, the hierarchy of higher equations and a dual equation for an asymmetric chiral f\/ield,
{\it Teoret. Mat. Fiz.} {\bf 70} (1987), 18--29
(English transl.: \href{http://dx.doi.org/10.1007/BF01017006}{{\it Theoret. and Math. Phys.}} {\bf 70} (1987), 11--19).

\bibitem{Orlov}
Orlov A.Yu.,
$N$-soliton solution of chiral f\/ields on Grassmann manifolds ($\sigma$ model),
{\it Teoret. Mat. Fiz.} {\bf 61} (1984), 214--225
(English transl.: \href{http://dx.doi.org/10.1007/BF01029111}{{\it Theoret. and Math. Phys.}} {\bf 61} (1984), 1099--1107).


\bibitem{We}
Weyl A.,
Elliptic functions according to Eisenstein and Kronecker,
{\it Ergebnisse der Mathematik und ihrer Grenzgebiete}, Band~88, Springer-Verlag, Berlin~-- New York, 1976.

\bibitem{Ma}
Mumford D.,
Tata lectures on theta.~I,
{\it Progress in Mathematics}, Vol.~28, Birkh\"auser Boston, Inc., Boston, MA, 1983.\\
Mumford D., Tata lectures on theta. II.~Jacobian theta functions and dif\/ferential equations,
{\it Progress in Mathematics}, Vol.~43, Birkh\"auser Boston, Inc., Boston, MA, 1984.

\end{thebibliography}
\end{document}